\documentclass[%
 reprint,
superscriptaddress,
 amsmath,amssymb,
 aps,
]{revtex4-1}

\usepackage[textsize=tiny]{todonotes}

\usepackage{amsbsy}
\usepackage{amsmath}
\usepackage{amsthm}

\usepackage{amssymb,subfigure}
\usepackage{braket}
\usepackage{color,xcolor}
\usepackage{graphicx}
\usepackage{bm}

\usepackage{cleveref}
\usepackage{subfigure}

\usepackage{comment}
\usepackage{epstopdf}
\usepackage{mathrsfs}

\newcommand{\ba}{\bm{a}}

\newcommand{\bc}{\bm{c}}

\newcommand{\be}{\bm{e}}
\newcommand{\bff}{\bm{f}}
\newcommand{\bg}{\bm{g}}

\newcommand{\bq}{\bm{q}}

\newcommand{\bu}{\bm{u}}
\newcommand{\bv}{\bm{v}}
\newcommand{\bw}{\bm{w}}
\newcommand{\bx}{\bm{x}}
\newcommand{\bz}{\bm{z}}

\newcommand{\bA}{\bm{A}}

\newcommand{\bD}{\bm{D}}
\newcommand{\bF}{\bm{F}}

\newcommand{\bI}{\bm{I}}
\newcommand{\bK}{\bm{K}}

\newcommand{\bM}{\bm{M}}

\newcommand{\bS}{\bm{S}}
\newcommand{\bT}{\bm{T}}
\newcommand{\bU}{\bm{U}}

\newcommand{\bzero}{\bm{0}}

\newcommand{\bmu}{\boldsymbol{\mu}}

\newtheorem*{prop*}{Proposition}
\newtheorem{prop}{Proposition}
\let\oldproposition\prop
\renewcommand{\prop}{\oldproposition\normalfont}

\newtheorem{thm}{Theorem}
\let\oldtheorem\thm
\renewcommand{\thm}{\oldtheorem\normalfont}

\newtheorem{defn}{Definition}
\let\olddefinition\defn
\renewcommand{\defn}{\olddefinition\normalfont}

\newcommand{\tred}[1]{{#1}} %%%%%%%%%% Uncomment this line for removing red color
\newcommand{\tblue}[1]{{#1}}

\begin{document}
\title{Amplitude-dependent topological edge states in nonlinear phononic lattices}

\author{Raj Kumar Pal}
\email{raj.pal@aerospace.gatech.edu}
\affiliation{School of Aerospace Engineering, Georgia Institute of Technology, Atlanta GA 30332}
\author{Javier Vila}
\affiliation{School of Aerospace Engineering, Georgia Institute of Technology, Atlanta GA 30332}
\author{Michael Leamy}
\affiliation{School of Mechanical Engineering, Georgia Institute of Technology, Atlanta GA 30332}
\author{Massimo Ruzzene}
\affiliation{School of Aerospace Engineering, Georgia Institute of Technology, Atlanta GA 30332}
\affiliation{School of Mechanical Engineering, Georgia Institute of Technology, Atlanta GA 30332}

%\author{Raj Kumar Pal$^{a,*}$, Javier Vila$^{a}$, Michael J. Leamy$^{b}$, Massimo Ruzzene$^{a,b}$ \\
%{\small $^a$ School of Aerospace Engineering, Georgia Institute of Technology, Atlanta GA 30332}\\
%{\small $^b$ School of Mechanical Engineering, Georgia Institute of Technology, Atlanta GA 30332}\\
%}

%\date{}

%\listoftodos

%\tableofcontents

%\tableofcontents

\begin{abstract}
This work investigates the effect of nonlinearities on topologically protected edge states in one and two-dimensional phononic lattices. We first show that localized modes arise at the interface between two spring-mass chains that are inverted copies of each other. Explicit expressions derived for the frequencies of the localized modes guide the study of the effect of cubic nonlinearities on the resonant characteristics of the interface which are shown to be described by a Duffing-like equation. Nonlinearities produce amplitude-dependent frequency shifts, which in the case of a softening nonlinearity cause the localized mode to migrate to the bulk spectrum.   
The case of a hexagonal lattice implementing a phononic analogue of a crystal exhibiting the quantum spin Hall effect is also investigated in the presence of weakly nonlinear cubic springs. An asymptotic analysis provides estimates of the amplitude dependence of the localized modes, while numerical simulations illustrate how the lattice  
response transitions from bulk-to-edge mode-dominated by varying the excitation amplitude. In contrast with the interface mode of the first example studies, this occurs both for  
hardening and softening springs.  The results of this study provide a theoretical framework for the investigation of nonlinear effects that induce and control topologically protected wave-modes through nonlinear interactions and amplitude tuning. 
\end{abstract}
\maketitle
\section{Introduction}

Wave propagation in periodic media is an active field of research with applications in diverse areas of science and engineering. 
Phononic crystals allow superior wave manipulation and control compared to conventional bulk media, 
since they present directional bandgaps and highly anisotropic dynamic behavior. Potential applications include vibration control, surface acoustic wave devices 
and wave steering~\cite{hussein2014dynamics}.  Recently, the achievement of defect-immune and scattering-free wave propagation using periodic media has received significant attention. The advent of topological mechanics~\cite{huber2016topological} provides an effective framework for the pursuit of 
robust wave propagation which is protected against perturbations and defects. 
Topologically protected edge wave propagation was originally envisioned in quantum systems and it has been 
quickly extended to other classical areas of physics, including acoustic~\cite{brendel2017snowflake}, photonic~\cite{khanikaev2013photonic}, 
optomechanical~\cite{peano2014topological} 
and elastic~\cite{mousavi2015topologically,pal2017edge} media. The unique properties achieved in these media, such as immunity to backscattering and localization in the presence of defects and imperfections, are a result of band topology. These properties allow for 
lossless propagation of information, or waves confined to a boundary or interface. Therefore, they may be part of a fundamentally new mechanism for wave-based transport of information or energy. 

There are two broad ways to realize topologically protected wave propagation in elastic media. The first one uses active components, thereby mimicking the quantum Hall effect. Changing of the parity of active devices or modulating of the physical properties in time have shown to alter the direction and nature of edge waves~\cite{swinteck2015bulk,nassar2017modulated}. Examples include magnetic fields in biological systems~\cite{prodan2009topological}, rotating disks~\cite{nash2015topological} and acoustic circulators operating on the basis of a flow-induced bias~\cite{khanikaev2015topologically}. The second way uses solely passive 
components and relies on establishing analogues of the quantum spin Hall effect. These media feature both forward and backward propagating
edge modes, which can be induced by an external excitation of appropriate polarization. The concept is illustrated in several studies by way of both 
numerical~\cite{mousavi2015topologically,pal2017edge,pal2016helical,he2016topological} and 
experimental~\cite{susstrunk2015observation,ningyuan2015time} investigations, which involve coupled pendulums~\cite{susstrunk2015observation}, 
plates with two scale holes~\cite{mousavi2015topologically} and resonators~\cite{pal2017edge}, 
as well as electric circuits~\cite{ningyuan2015time}.  Numerous studies have also been conducted on localized non-propagating 
deformation modes at the interface of two structural lattices~\cite{prodan2017dynamical,pal2017edge,chaunsali2017demonstrating}. 
These modes depend on the topological properties of the bands, and in $1D$ lattices, they are characterized by the Zak phase as the topological invariant~\cite{xiao2015geometric}. In 
$2D$ and $3D$ lattices, several researchers have investigated the presence of floppy modes of motion due to nontrivial 
topological polarization and exploited these modes to achieve localized buckling and directional response~\cite{kane2014topological,paulose2015selective,rocklin2016mechanical,rocklin2016directional}.  

While most studies consider systems governed by linear interactions, there is growing interest in the investigation of the effect of nonlinearities in topological 
materials. Nonlinearities, for example, enable tunable wave motion, which in turn may lead to 
non-reciprocal wave propagation~\cite{fleury2014sound,vila2017bloch}. This finds potential applications in acoustic switching~\cite{pal2014wave}, 
diodes~\cite{boechler2011bifurcation} and delay lines~\cite{alu2016metamaterials}. Nonlinear effects have been investigated to demonstrate self-induced 
topological phase transitions in SSH (Su-Schrieffer-Heeger)~\cite{hadad2016self}. In the field of photonics, several studies have considered topological effects in nonlinear media.  Included in these studies are soliton-like topological states which exist on the edges of weakly nonlinear photonic systems \cite{ablowitz2014linear,leykam2016edge}, or in the bulk and propagate around the edge of a self-induced defect \cite{lumer2013self}.  These solitons arise in systems that can be described by a nonlinear Schrodinger (NLS) equation \cite{ablowitz2014linear,ablowitz2017tight,lumer2013self,leykam2016edge} and coupled nonlinear SSH equations \cite{zhou2017optical}, all obtained from a Kerr-like optical nonlinearity.

%A common feature of nonlinear wave motion, is its amplitude dependent properties and response, which provide the means for affecting, or ``tuning,'' the behavior of the system by varying the excitation amplitude. These investigations can take advantage of perturbation methods for the study of wave motion in nonlinear periodic media, as presented for the case of weakly nonlinear interaction in $1D$ and $2D$ lattices~\cite{narisetti2010perturbation,Narisetti2011} 

This work investigates the effect of nonlinearities on two types of topologically protected localized modes in phononic lattices. Specifically, we study the robustness and frequency content of localized modes in a $1D$ and $2D$ lattices. In the $1D$ lattice, we illustrate the amplitude-dependent resonant behavior of an interface mode, which can lead to its shifting into the bulk bands.  In the $2D$ case, the perturbation approach of  ~\cite{narisetti2010perturbation,Narisetti2011} is applied to predict the amplitude-dependent frequency of edge modes for both hardening and softening springs. In both cases, the predictions are verified through numerical simulations on finite lattices excited by forces of increasing amplitude. 

The outline of this paper is as follows: Sec.~\ref{sec:1Dlattice} presents a discrete $1D$ lattice (chain) with an interface 
and explicit expressions for the frequency and mode shapes of modes localized at the interface. 
The corresponding tunable nonlinear chain version is discussed in Sec.~\ref{sec:1Dnonlinear}. 
Then we show how the lattice response can switch from bulk to  edge waves at a fixed frequency 
by varying amplitude in a $2D$ lattice in Sec.~\ref{sec:2Dlattice}. The $2D$ designs are 
verified by a combination of dispersion analysis and numerical simulations on finite lattices.
Finally, Sec.~\ref{sec:concl} presents the conclusions of this study.

%%%%%%%%%%%%%%%%%%%%%%%%%%%%%%%%%%%%%%%%%%%%%%%%%%%%%%%%%%%%%%%%%%%%%%%%%%%%%%%%%%%%%%%%%%%%%%%%%%%%%%%%%%%%%%

\section{Interface modes in a $1D$ lattice}\label{sec:1Dlattice}

We begin our investigations by illustrating the existence and behavior of interface modes in a $1D$ spring-mass chain. The linear case is presented first, in order to briefly describe
the existence of localized modes at the interface of chains that are characterized by distinct topological invariants, which in this case is the Zak phase~\cite{xiao2015geometric,xiao2014surface}. The analytical derivation of the interface mode frequencies are presented in Appendix~\ref{Sec.TransferMatrix} and details of the topological properties of the linear chain are provided in Appendix~\ref{Sec.BandInversion}. Next, the behavior of an interface with nonlinear interactions is investigated in detail through its representation as a simple, single degree of freedom oscillator. This approach enables the study of the effect of nonlinearities in relation to the existence of the interface mode as a function of the excitation amplitude, and specifically to its tendency to enter the bulk spectrum based on the parameters defining the nonlinear interactions.

\subsection{Linear chain: Analytical and numerical results}\label{sec:1Danalytic_edgemode}

\begin{figure}[hbtp]
	\centering
	\includegraphics[width=0.5\textwidth]{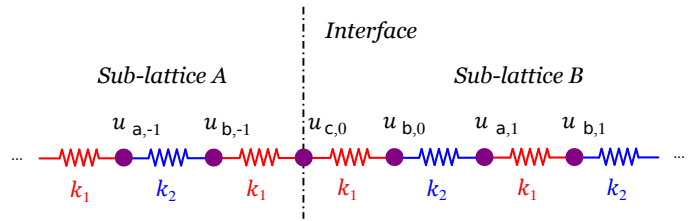}
	\caption{Two sub-lattices (A,B) which are inverted copies of each other, are joined together. The interface supports  
	support a localized mode in the bandgap frequencies.}
	\label{Fig.InterfaceFiniteSystem}
\end{figure}
The spring mass chain model considered is 
displayed in Fig. \ref{Fig.InterfaceFiniteSystem}. It consists of two sub-lattices, each with identical masses and having alternating springs with
stiffness $k_1$ and $k_2$ and of an  interface (or defect) mass connecting them. The
interface mass is connected to springs with stiffness $k_1$ on both sides. The unit cells on the right and left of this interface are inverted copies of each other. 
This discrete lattice was investigated in~\cite{pal2017edge}, where the existence of two types of localized modes were 
discussed. 

The governing equation for the free vibration of interface mass is 
\begin{equation}
m \ddot{u}_{c,0} + k_1 \left(2u_{c,0} - u_{b,0} - u_{b,1} \right)  = 0 \\
\end{equation}
Similarly, the governing equations for a unit cell $p$ of the sub-lattice on the left of the interface
\begin{subequations}\label{Eqn_left}
\begin{gather}
m \ddot{u}_{a,p} + k_2 \left(u_{a,p} - u_{b,p} \right) + k_1 \left(u_{a,p} - u_{b,p-1} \right) = 0 \\
m \ddot{u}_{b,p} + k_2 \left(u_{b,p} - u_{a,p} \right) + k_1 \left(u_{b,p} - u_{a,p+1} \right) = 0 
\end{gather}
\end{subequations}
while for a unit cell $p$ on the right sub-lattice are
\begin{subequations}\label{Eqn_right}
\begin{gather}
m \ddot{u}_{a,p} + k_1 \left(u_{a,p} - u_{b,p} \right) + k_2 \left(u_{a,p} - u_{b,p-1} \right) = 0 \\
m \ddot{u}_{b,p} + k_1 \left(u_{b,p} - u_{a,p} \right) + k_2 \left(u_{b,p} - u_{a,p+1} \right) = 0. 
\end{gather}
\end{subequations}
The above equations are normalized by writing the spring constants as $k_1 = k(1+\gamma)$ and 
$k_2 = k(1-\gamma)$, with $\gamma$ being a stiffness parameter and $k$ being the mean stiffness. A nondimensional time scale
$\tau = \sqrt{k/m}t$ is introduced to express the equations in nondimensional form. 

In the present work, explicit expressions for the frequency of the localized modes at the interface are derived by using a transfer matrix 
approach. The derivations can be found in Appendix~\ref{Sec.TransferMatrix}. These expressions allow us to identify and investigate the parameters affecting the frequency and modeshapes in a systematic 
way. 
\tred{ They also shed light on the amplitude dependence (for $\gamma < 0$) and independence (for $\gamma > 0$) of the localized modes in 
chains with weakly nonlinear springs. }
Our theoretical predictions are verified through a combination of frequency domain analysis and transient numerical simulations on a finite chain with an interface. 

The following are the solutions for the frequencies which support localized solutions
\begin{equation}\label{Eqn.xtra3.Appendix}
\begin{split}
& \gamma < 0:\quad\Omega=\sqrt{ 3- \sqrt{1 + 8 \gamma^2}},\quad\textrm{Anti-symmetric mode} \\
& \gamma > 0:\quad\Omega=\sqrt{2},\;\;\qquad\qquad\quad\quad\textrm{Symmetric mode} \\
& \gamma > 0 :\quad\Omega=\sqrt{ 3+ \sqrt{1 + 8 \gamma^2}},\quad\textrm{Anti-symmetric mode. } 
\end{split}
\end{equation}
\tblue{Here $\Omega$ is the non-dimensional frequency obtained by normalizing with the reference frequency $\sqrt{k/m}$. }
The detailed derivations of these frequencies along with their associated modeshapes are presented
in appendix~\ref{Sec.TransferMatrix}.
Note that the first and second solutions give frequencies which are localized in the bandgap between the acoustic and optical branches, while the third frequency is above the optical branch. Furthermore, the first and third frequencies are  
associated with anti-symmetric mode shapes where the unit cells on both sides of the interface are in phase, while the second frequency is
associated with a symmetric mode shape, with the interface mass being at rest, while the unit cells on both sides have a phase difference of $\pi$.
%Thus the three frequencies presented above correspond to all the possible interface frequencies observed for both kinds of interface.
\begin{figure}[hbtp]
	\centering
	\subfigure[]{
	\includegraphics[width=0.32\textwidth]{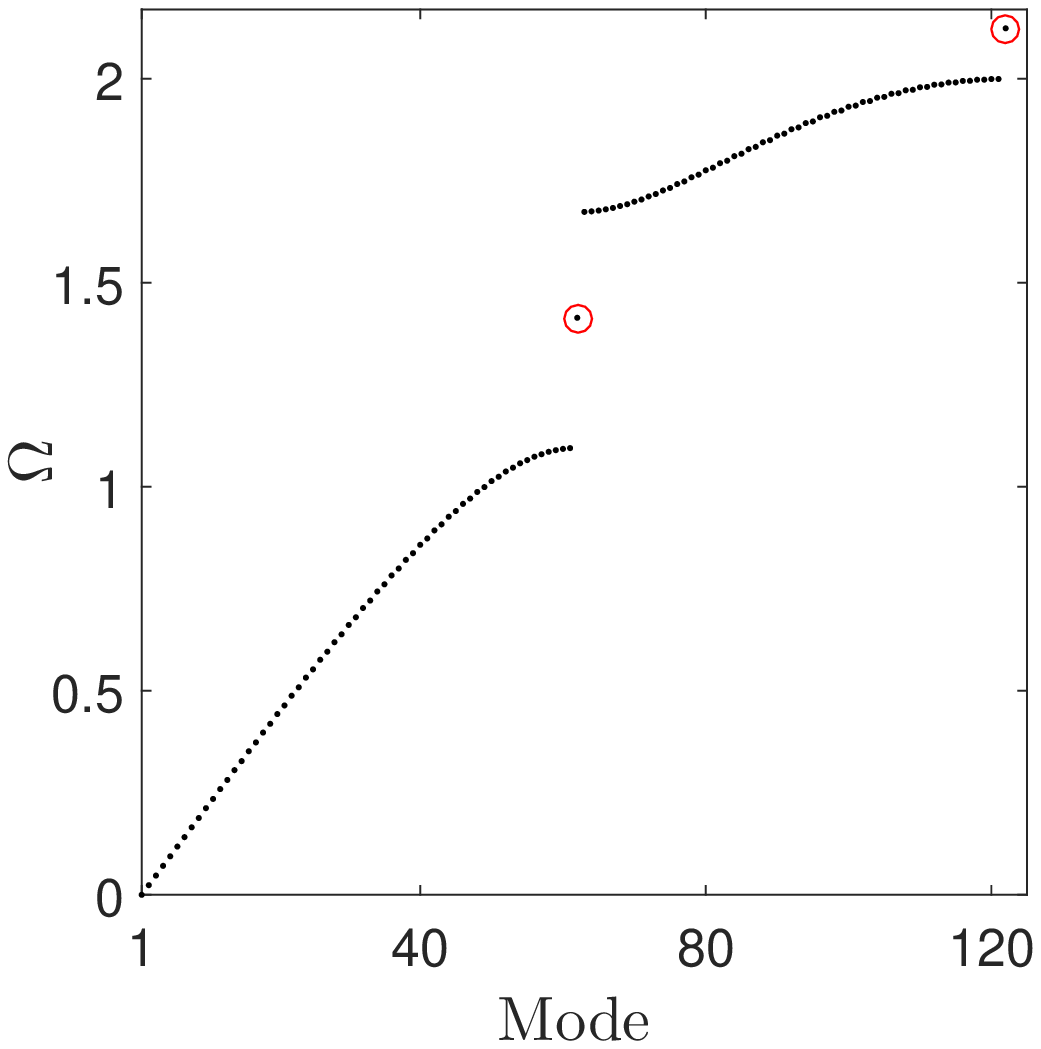}
	\label{Fig.LinearModes}}
\subfigure[]{
	\includegraphics[width=0.32\textwidth]{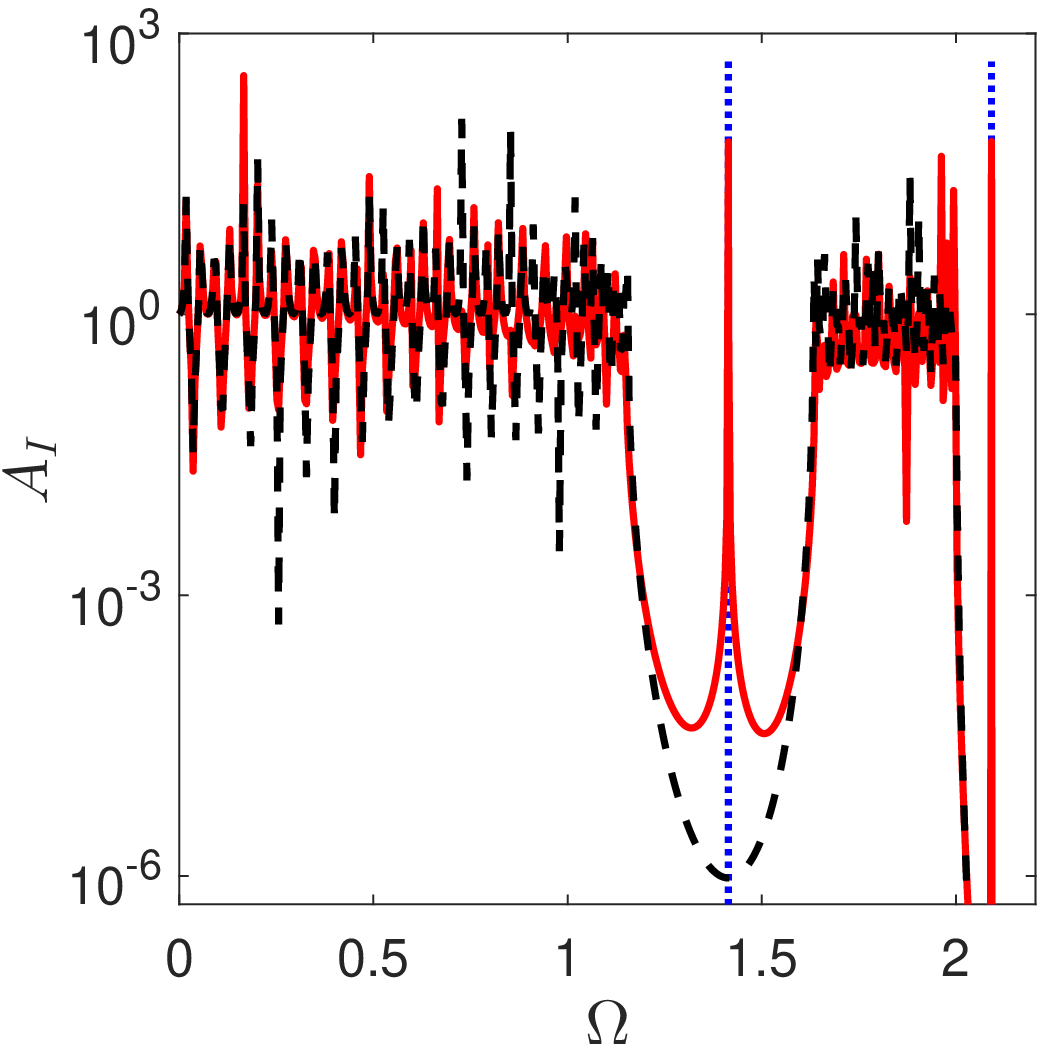}	
	\label{Fig.FRFLineal}
}
	\caption{ (a) Natural frequencies of a finite chain exhibiting 
	interface modes (red circles) in the bandgap frequencies. 
	(b) Frequency response function (red,solid) showing the interface mode within the bandgap, in agreement with 
analytical predictions (blue, dotted). Interface modes are absent in a regular chain with all identical unit cells (black, dashed). 
	}
\end{figure}

We verify our analytical predictions by numerically computing 
the modes of oscillation of a finite chain having $60$ unit cells with an interface at the center, see
Figure~\ref{Fig.InterfaceFiniteSystem}. The stiffness parameter is set to $\gamma=0.4$. 
The governing equations for our lattice may be written in matrix form as $\bM \ddot {\bq}(\tau) + \bK \bq(\tau) = \bff(\tau)$. 
We seek the forced vibration response of the linear chain when subjected to an external force $\bff \cos\Omega \tau$. Imposing a solution ansatz of the form $\bq(\tau) = \bq e^{i\Omega\tau}$, 
the governing equation reduces to 
\begin{equation}\label{govE.Linear}
\left( \bK - \Omega^2 \bM\right) \bq = \bff . 
\end{equation}
Figure~\ref{Fig.LinearModes} displays the natural frequencies $\Omega$ of this chain, obtained by solving the eigenvalue problem that arises by setting
$\bff = 0$. It illustrates the presence of a bandgap between the 
acoustic and optical modes. 
Furthermore, there is an interface mode in the bandgap at frequency $\Omega = \sqrt{2}$, which matches exactly with 
the analytical solution of $\Omega_i$ for $\gamma > 0$ in Eqn.~\eqref{Eqn.xtra3.Appendix}. Analogous results are obtained
for the chain with $\gamma<0$, consistent with the analytical expressions for the localized mode frequencies and shapes. 

To illustrate the dynamic behavior of this chain, we compute the frequency response function by imposing a displacement $ u_{b,30}=\cos(\Omega\tau)$ on the mass at the left boundary. 
The other end of the chain is free and the frequency response is normalized with the excitation amplitude, which is unity in our study. 
We also consider a chain that has no interface and comprises $60$ 
identical unit cells (regular chain), in which edge modes are not expected. 
Figure~\ref{Fig.FRFLineal} displays the displacement amplitude of the center mass for both chains obtained by solving Eqn.~\eqref{govE.Linear} with appropriate displacement boundary conditions $(\bff = 0)$ over a wide frequency range. In the bandgap frequency range, the regular chain with all identical unit cells does not support any resonance mode. The chain with an interface mass has a resonance mode, consistent with the analytical solution (Eqn. \eqref{Eqn.xtra3.Appendix}).

\subsubsection{Reduced model for forced response}\label{Sec.LinearReducedModel}
We now seek the forced vibration response of a chain comprised of $N$ unit cells on each side of the interface. The interface mass
is subjected to an external forcing $f$ at frequency $\Omega$. We consider the anti-symmetric mode that arises when $\gamma < 0$, for which we derive a reduced order model when the 
interface mass is subjected to the external force. As shown in Eqn.~\eqref{Eqn.xtra3.Appendix}, the interface 
mode is anti-symmetric, i.e., $u_{b,0} = u_{b,-1}$. 
Since the wavenumber is $\pi$ in the bandgaps 
and there is no propagation, this displacement relation is valid for frequencies in the bandgap when $\gamma < 0$. 
The relation $\bu_N = \bT^N \bu_0$ can be inverted to get the relation
$\bu_0 = \bT^{-N} \bu_N$. 
We reduce the chain to a single degree of freedom system which governs the behavior of the interface mass and  
obtain an expression for the effective stiffness 
on the interface mass. 
Fixing the first and last masses of the chain ($u_{b,N} = u_{b,-N}= 0$), 
the relation $\bu_0 = \bT^{-N} \bu_N$ simplifies to the equation $u_{b,0} / S_{12} = u_{c,0}/S_{11}$, where $S_{ij}$ are the components
of $\bS =\bT^{-N}$. The governing equation of the interface mass is $(2(1+\gamma)-\Omega^2)u_{c,0} - 2(1+\gamma)u_{b,0} = f$. 
Eliminating $u_{b,0}$ from these
two relations yields the following expression for the effective behavior of the interface mass 
\begin{equation}\label{Eqn.LinearReduced}
\left[ 2(1+\gamma)\left( 1-\dfrac{S_{21}}{S_{11}} \right)  - \Omega^2 \right] u_{c,0}  = f. 
\end{equation}
Explicit expressions for the terms $S_{11}$ and $S_{21}$ in terms of the excitation frequency $\Omega$ and $\gamma$ are presented in 
Appendix~\ref{Sec:ReducedSoln}.

%\subsubsection{Numerical simulations on a linear lattice}
%
%\begin{figure}[hbtp]
%	\centering
%	\includegraphics[width=0.32\textwidth]{4c.eps}
%	\caption{ Frequency response in the bandgap frequencies along with the forced vibration response (transient solution, dot-dashed line) of 
%	the reduced model. Excellent agreement is observed. 
%	}
%	\label{Fig.TmatchA}
%\end{figure}
%
%Figure~\ref{Fig.TmatchA} displays the frequency response in the bandgap frequencies along with the 
%response obtained from transient simulations of the reduced model. We note here 
%that transient simulations for a finite lattice also yield the same frequency response as the reduced model within the bandgap. The dynamic response 
%shows a high amplitude when excited at the interface mode frequency, thereby illustrating a good agreement with the theoretical 
%predictions and frequency domain analysis.  

\begin{figure}[hbtp]
	\centering
\subfigure[]{
		\includegraphics[width=0.52\textwidth]{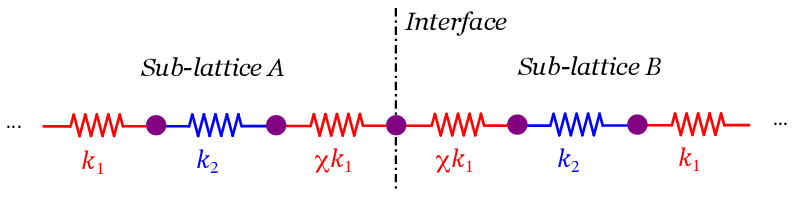}
	\label{Fig.3a}
}
\subfigure[]{
		\includegraphics[width=0.32\textwidth]{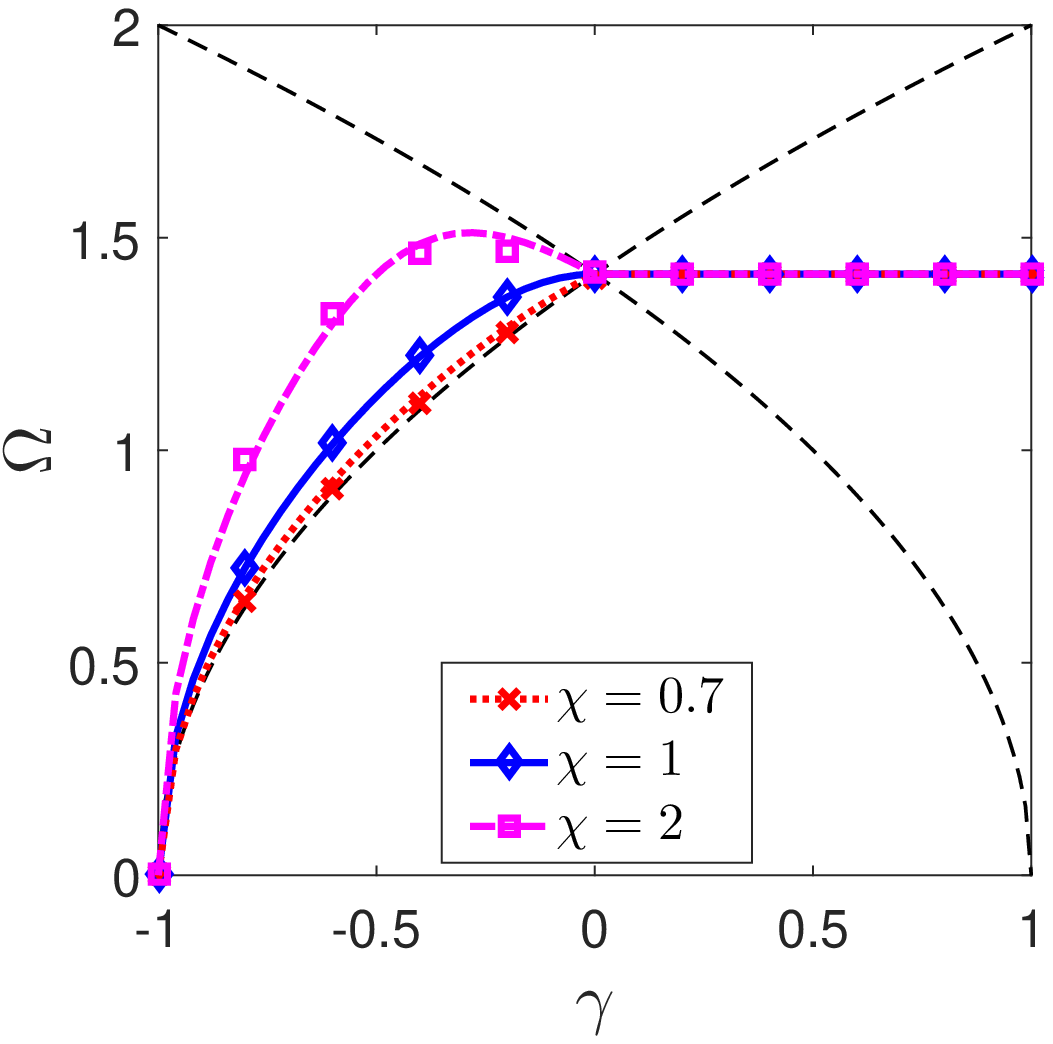}
	\label{Fig.xtra1000}
}
	\caption{(a) Schematic of an interface having springs with modified stiffness $\chi k_1$. All other spring stiffnesses remain unchanged. 
(b) \tblue{Variation of the interface frequency as a function of $\gamma$ for the $3$ distinct values of $\chi$ listed in the legend. } Dashed curves show frequencies bounding the bandgap.}
\end{figure}

The results for the interface frequency can be further generalized to the case of 
springs adjacent to the interface different from $k_1$ through a parameter $\chi$. 
The springs connected to the interface mass are changed to $\chi k_1$ while the stiffness of all the other 
springs in the chain remain unchanged. Figure~\ref{Fig.3a} illustrates a schematic of this modified interface.  
The governing equation for the interface mass now becomes 
\begin{equation*}
-\Omega^2 u_{c,0} + \chi(1+\gamma) (2u_{c,0} - u_{b,0} + u_{b,-1}) = 0. 
\end{equation*}
Numerical analysis is performed to determine the natural frequencies of this modified interface using 
a chain of $60$ unit cells by solving the eigenvalue problem 
($\bff = \bzero$ in Eqn.~\eqref{govE.Linear}). The interface mode frequency is located by examining its corresponding mode shape. 
\tblue{Figure~\ref{Fig.xtra1000} displays the interface mode frequency for three distinct $\chi$ values: $0.7$, $1$ and $2$ over a range of stiffness parameter values $\gamma$.} 
Also shown by dashed lines are the frequencies bounding the bandgap. 
We observe that only the anti-symmetric mode ($\gamma<0$) frequency shifts, while the symmetric mode ($\gamma> 0$) frequency 
remains unchanged.  
This surprising observation can be explained by examining the mode shape of the edge mode when $\gamma > 0$. 
For this mode shape, the interface mass $(c,0)$ has zero displacement as the force acting on its either sides are equal 
and opposite. Thus changing the stiffness of the spring connecting $(c,0)$ from both sides does not affect the dynamic behavior of this mass. The 
displacement of the adjacent masses $u_{b,0}$ and $u_{b,-1}$ change in account of the increased stiffness. However the remaining
mode shape and the corresponding frequency do not change with $\chi$.

\subsection{Analysis of nonlinear interface}\label{sec:1Dnonlinear}
Based on the above observations, we seek to achieve a tunable response in our chain by using nonlinear springs whose stiffness 
depends on the amplitude. An effect similar to the frequency shift due to springs with stiffness $\chi k_1$ in the 
above linear chain may be obtained by varying the excitation force amplitude. 
We consider a chain identical to the above linear chain with an interface, 
but replace the two interface springs having stiffness $\chi k_1$ with weakly nonlinear springs, whose restoring 
force varies with relative displacement $\Delta u$ as
$F=k_1 \Delta u + \Gamma(\Delta u)^3$. 
Adding a cubic nonlinearity leads to an amplitude-dependent frequency of the interface mode. 
Viscous damping with coefficient $c$ is applied to the mass at the interface so that a steady-state can be reached
in our numerical simulations.
We show how the nonlinear chain behaves essentially as a Duffing oscillator using a reduced order model for the interface mass, similar to 
Eqn.~\eqref{Eqn.LinearReduced}. \tred{Our analytical results thus provide the opportunity to apply 
known results on Duffing oscillators to the investigation of edge modes in nonlinear regimes. }

We investigate the forced vibration response of this nonlinear chain subjected to an external excitation force $f$ 
applied at the interface mass. 
The governing equation for the mass at the interface and its adjacent masses may be written as 

\begin{align}
&m\ddot{u}_{c,0} +  c \dot{u}_{c,0} + k_1(2u_{c,0} - u_{b,0} - u_{b,-1})  + \nonumber\\
					&\qquad \Gamma \left( u_{c,0} - u_{b,0}  \right)^3 
					+ \Gamma\left(  u_{c,0} - u_{b,-1}   \right)^3 = f\cos (\Omega t),  \nonumber \\ 
&m\ddot{u}_{b,0} +  k_1\left(u_{b,0} - u_{c,0}\right) + k_2 \left(u_{b,0} - u_{a,1} \right) + \nonumber  \\
					&\;\;\qquad\qquad\qquad\qquad\qquad  \Gamma ( u_{b,0} - u_{c,0}  )^3 = 0 ,  \\ 
&m\ddot{u}_{b,-1} +  k_1\left(u_{b,-1} - u_{c,0}\right) + k_2 \left(u_{b,-1} - u_{a,-1} \right) + \nonumber   \\
					&\qquad\qquad\qquad\qquad\qquad  \Gamma ( u_{b,-1} - u_{c,0}  )^3 = 0. \nonumber 
\end{align}
The governing 
equations for all the other masses on both sides of the interface remain the same as in the linear case 
(Eqns.~\eqref{Eqn_left},~\eqref{Eqn_right}). 
Again, we consider a chain with stiffness parameter $\gamma < 0$ and derive the equivalent behavior of the interface mass in the bandgap frequencies. 

To now get an equivalent equation for the interface mass, we need 
to eliminate $u_{b,0}$ from the governing equation of the interface mass.  
Let us assume an approximate solution for the displacement of the masses in the chain to be of the form 
\begin{equation}\label{eqn_HB}
\bu = \dfrac{\bv e^{i\Omega t}}{2} + \epsilon \sum_{n = 2}^M \left( \bw_n e^{i n \Omega \tau} \right) + c.c. ,
\end{equation}
with $\epsilon$ being a bookkeeping parameter and $c.c.$ denoting the complex conjugate. 
Recall that $e_1$ and $e_2$ are the components of the 
eigenvector corresponding to the localized mode in the linear chain (Eqn.~\eqref{Eqn.Sol}). 
The nonlinear force term may be approximated as
\begin{equation}\label{approx1}
\Gamma (u_{c,0}  - u_{b,0})^3 = \dfrac{3}{8}\Gamma \left(1 - \dfrac{e_2}{e_1}\right)^3 |v_{c,0}|^2 v_{c,0} e^{i\Omega t} + \epsilon(h.h.),
\end{equation}
where $h.h.$ denotes higher harmonics. 
Note that the above approximation is valid for small displacements when the term $u_{b,0}/u_{c,0}$ can be approximated by 
the linear solution ($\bu_0 = s\be$). 

We perform a harmonic balance on the linear parts of the chain 
by considering only the terms of frequency $\Omega$.  
The displacements in the linear parts of the chain can be related using the transfer matrix approach. 
Observe that the structure of the chain results in exactly the same relation as Eqn.~\eqref{Eqn.TransferMatrix} 
holding between $(v_{b,p-1},v_{a,p})$ and $(v_{b,p},v_{a,p+1})$
under the transformation $\gamma \to -\gamma$. 
Thus, defining the corresponding 
quantities $\overline{\bS}(\gamma) = \bS(-\gamma) = \bT^{-N}(-\gamma)$ leads to the following relation
\begin{equation}\label{barSeqn}
\overline{S}_{12}v_{b,0}  -  \overline{S}_{11} v_{a,1}  =  0. 
\end{equation}
Imposing Eqn.~\eqref{eqn_HB} and again performing a harmonic balance, the equation for the displacement $v_{b,0}$ of the 
mass adjacent to the interface mass now becomes 
\begin{multline*}
-\Omega^2 v_{b,0} +  (1+\gamma)\left(v_{b,0} - v_{c,0}\right) + (1-\gamma) \left(v_{b,0} - v_{a,1} \right) \\
					- \dfrac{3\Gamma}{4} \left( 1 - \dfrac{e_2}{e_1}  \right)^3 |v_{c,0}|^2 v_{c,0} = 0, 
\end{multline*}
Eliminating $v_{a,1}$ from the above equation using Eqn.~\eqref{barSeqn}, it may be rewritten as
\begin{multline}\label{Eqn_ub0}
\left[ 2 - \Omega^2 - (1-\gamma)\dfrac{\overline{S}_{12}}{\overline{S}_{11}} \right]v_{b,0}  = \\
\dfrac{3\Gamma}{4} \left( 1 - \dfrac{e_2}{e_1}  \right)^3 |v_{c,0}|^2 v_{c,0} + (1+\gamma)v_{c,0} . 
\end{multline}

We may write an equation similar to Eqn.~\eqref{Eqn_ub0} for the displacement $u_{b,-1}$ of the mass at the left of the interface mass
and use an approximation similar to Eqn.~\eqref{approx1} to simplify its cubic nonlinear term. Indeed, for the case $\gamma < 0$, recall that 
the zeroth order solution is an anti-symmetric mode and thus $v_{b,0} = v_{b,-1}$. 
Substituting Eqn.~\eqref{Eqn_ub0} and its counterpart for $u_{b,-1}$ 
into the governing equation for the interface mass and performing 
a harmonic balance again leads to the following equation
\begin{multline}
-\Omega^2{v}_{c,0} +  i \Omega \delta v_{c,0} + 
(1+\gamma)\left( 2 - \dfrac{1+\gamma}{g}\right)  v_{c,0} 
+ \\  \dfrac{3\Gamma}{4} \left(1- \dfrac{1+\gamma}{g}\right)\left(1 - \dfrac{e_2}{e_1}\right)^3 |v_{c,0}|^2 v_{c,0} = f,
\end{multline}
where $g =2-\Omega^2-(1-\gamma)\overline{S}_{12}/\overline{S}_{11} $ and 
$\delta = c \sqrt{2/k m}$ is the nondimensional damping parameter. 
Decomposing $v_{0,c} = v_R + iv_I$ into its real and imaginary parts leads to two equations. Squaring and summing them 
leads to the following frequency amplitude response~\cite{nayfeh2008perturbation}
\begin{equation}\label{DuffingFinal}
\left[ \left( \Omega^2 - k_{e}- \dfrac{3}{4}\Gamma_{e} u^2\right)^2 + (\delta\Omega)^2\right]u^2 = f ^2, 
\end{equation}
with $u = |v_{0,c}|$ being the displacement amplitude and 
\begin{gather*}
k_{e} = (1+\gamma)\left( 2 - \dfrac{1+\gamma}{g}\right) , \\
\Gamma_{e} = \Gamma \left(1- \dfrac{1+\gamma}{g}\right)\left(1 - \dfrac{e_2}{e_1}\right)^3 . 
\end{gather*}
The above frequency amplitude response is similar to that of a Duffing oscillator with linear stiffness $k_{e}$ and nonlinear
force $\Gamma_{e}$ excited near the resonant frequency~\cite{nayfeh2008perturbation}. 

Let us first consider  a chain with strain hardening springs ($\Gamma > 0$) connected to the interface mass. 
The dynamic response of the chain is investigated using the amplitude-response predicted by the reduced order model (Eqn.~\eqref{DuffingFinal}) and transient simulations of the full nonlinear chain, performed using the Verlet algorithm \cite{Verlet67}.
We compare the frequency response function predicted by the reduced order model with numerical simulations on 
a finite chain. The numerical simulations are performed until the chain attains a steady state. 
The damping coefficient, linear and nonlinear stiffness parameter values are set to 
$\delta = 0.01$, $\gamma = 0.4$ and $\Gamma = 0.1$, respectively. The interface mass is subjected to an external force $f\cos(\Omega\tau)$. 
The frequency response is computed by normalizing the displacement $u_{c,0}$ of the 
interface mass by the excitation force amplitude $f$ as $A_I=u_{c,0} k / f$. 

\begin{figure}[hbtp]
	\centering
	\subfigure[]
	{\includegraphics[width=0.32\textwidth]{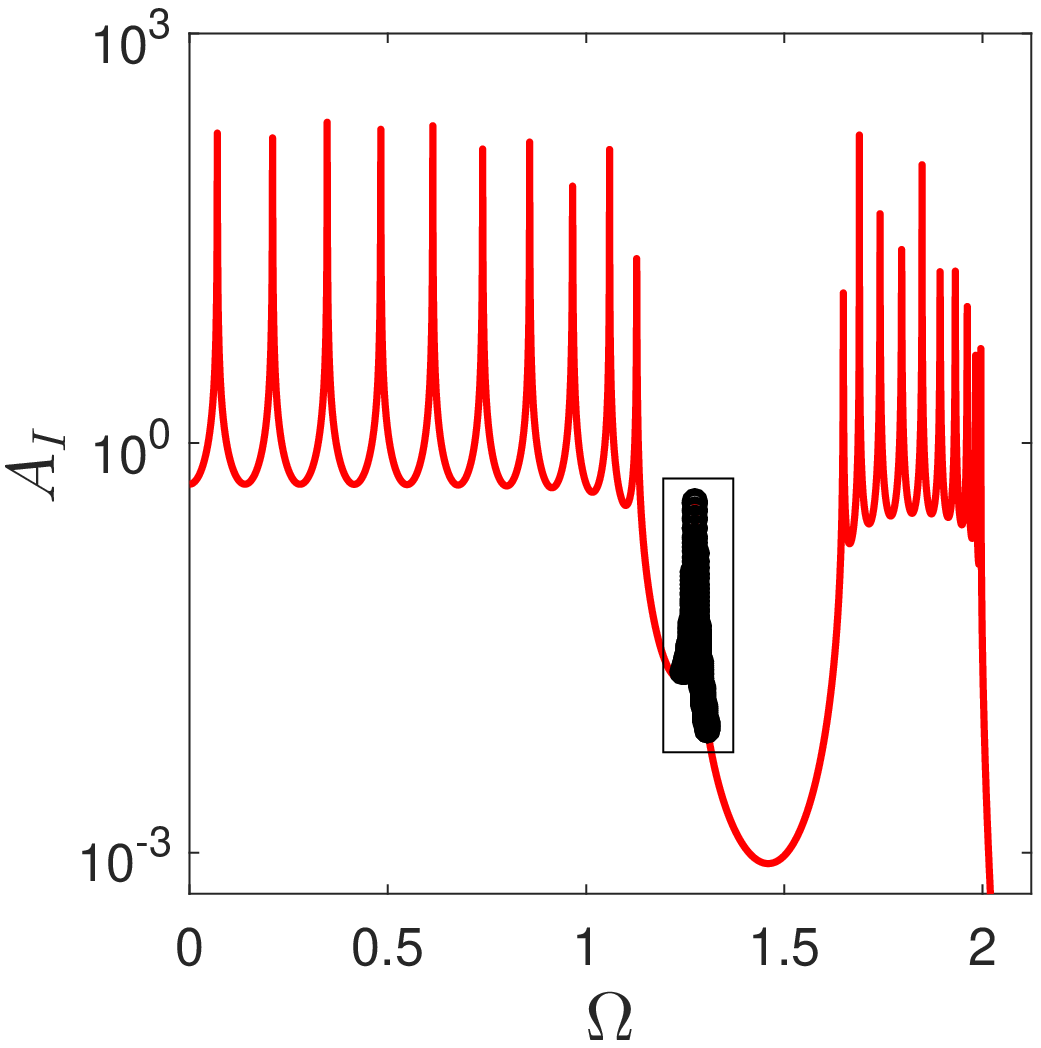}
	\label{Fig.FRF_full_1}}
	\subfigure[]
	{\includegraphics[width=0.32\textwidth]{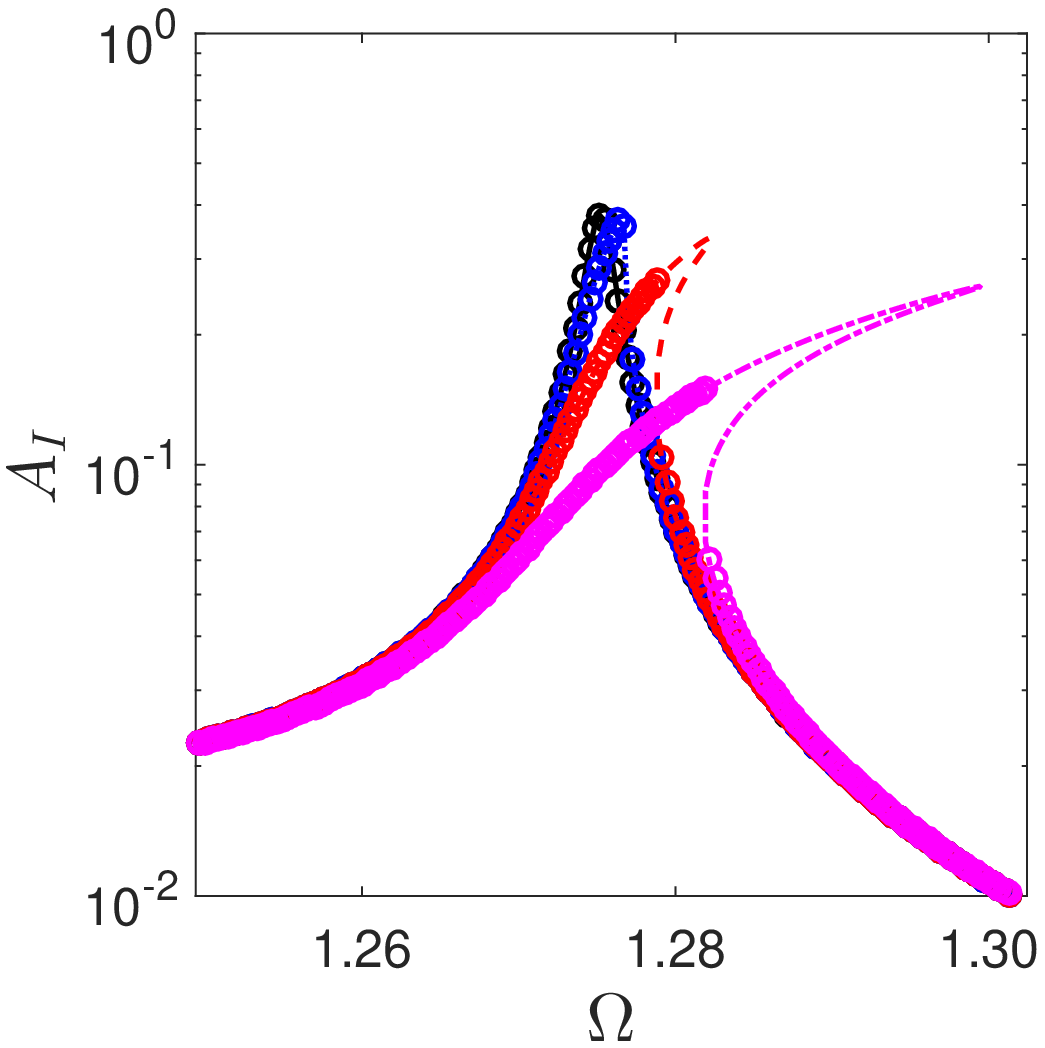}
	\label{Fig.FRF_close_1}}
	\caption{Frequency response of the interface mass normalized by excitation force $f$. (a) Both 
	finite chain numerical simulations (red curve) and analytical solution of reduced model (black circles) show an interface mode 
	for small forcing amplitude $f=1$. (b) Numerical (markers) and analytical (curves) solutions for various 
	force amplitudes $f=\{1,4,10,25\}$.
	Curves shift to the right and the chain behaves as a Duffing oscillator for frequencies near the interface mode. }
	\label{Fig.FRFNoLineal2}
\end{figure}

Figure~\ref{Fig.FRF_full_1} displays the frequency response of a finite linear chain (red curve) over $\Omega \in [0,\; 2]$ along with the response observed from simulations of the finite nonlinear chain (black circles) for frequencies in the vicinity of the interface mode frequency, when subjected to low amplitude excitation ($f_0=1$). 
The linear chain response is obtained by solving the forced vibration response at steady state using Eqn.~\eqref{govE.Linear}. 
Since the excitation force amplitude 
is low, nonlinear effects are seen to be negligible, and the predictions of the linear model are in good
agreement with the numerical simulations for frequencies near the interface mode frequency.
 
Figure~\ref{Fig.FRF_close_1} displays a close up view of the frequency response computed 
from simulations of the finite nonlinear chain (markers), along with the response given by Eqn.~\eqref{DuffingFinal}, 
the nonlinear reduced order model (solid curves),
for various force excitation amplitudes $f = \{1,4,10,25\}$. 
An excellent agreement is obtained between them which confirms the validity of our reduced order model. 
The peak force shifts to the right with
increasing force amplitude and displays a backbone curve. This behavior is typical of a Duffing oscillator~\cite{nayfeh2008perturbation} 
and demonstrates the amplitude dependent behavior 
of the interface mode. 

\begin{figure*}[hbtp]
	\centering
	\subfigure[]{
		\includegraphics[width=0.32\textwidth]{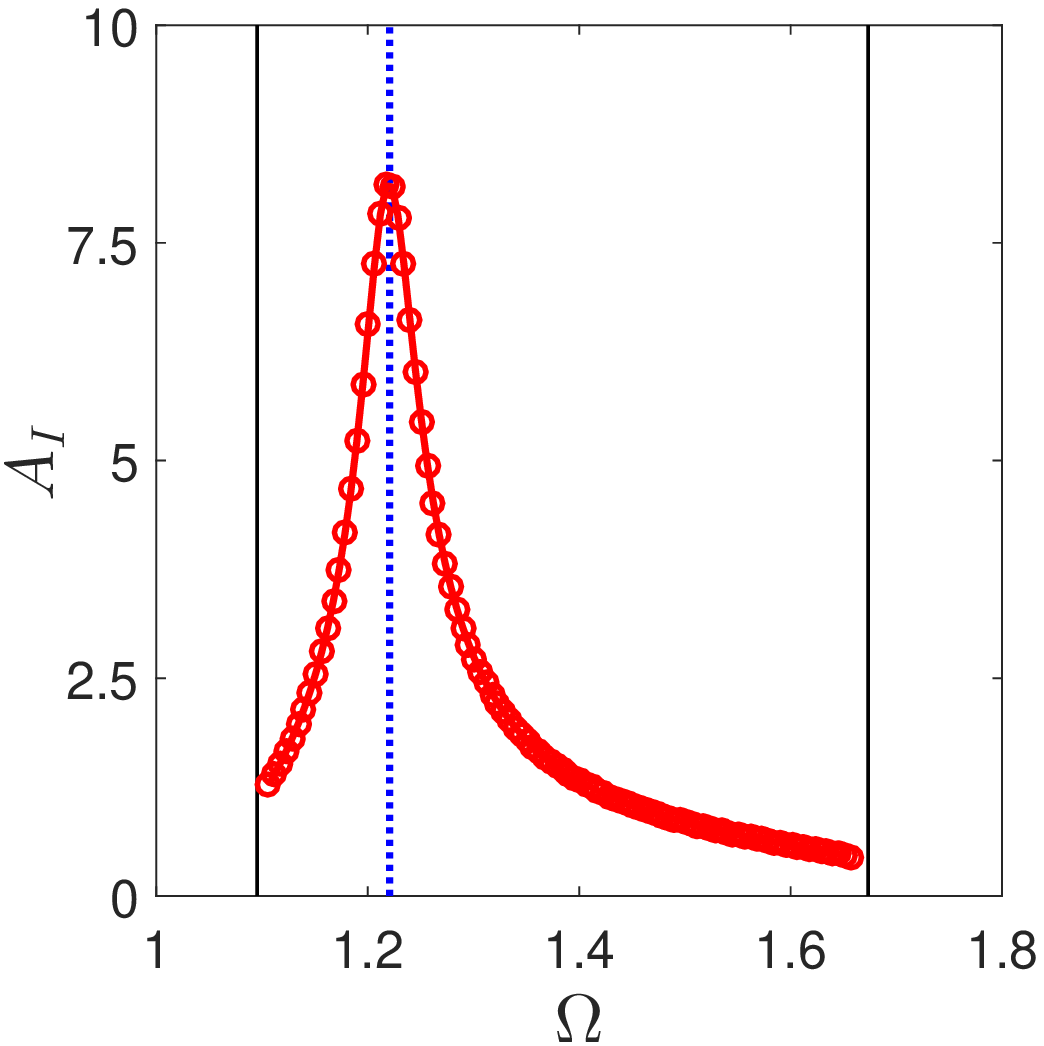}}
	\subfigure[]{
		\includegraphics[width=0.32\textwidth]{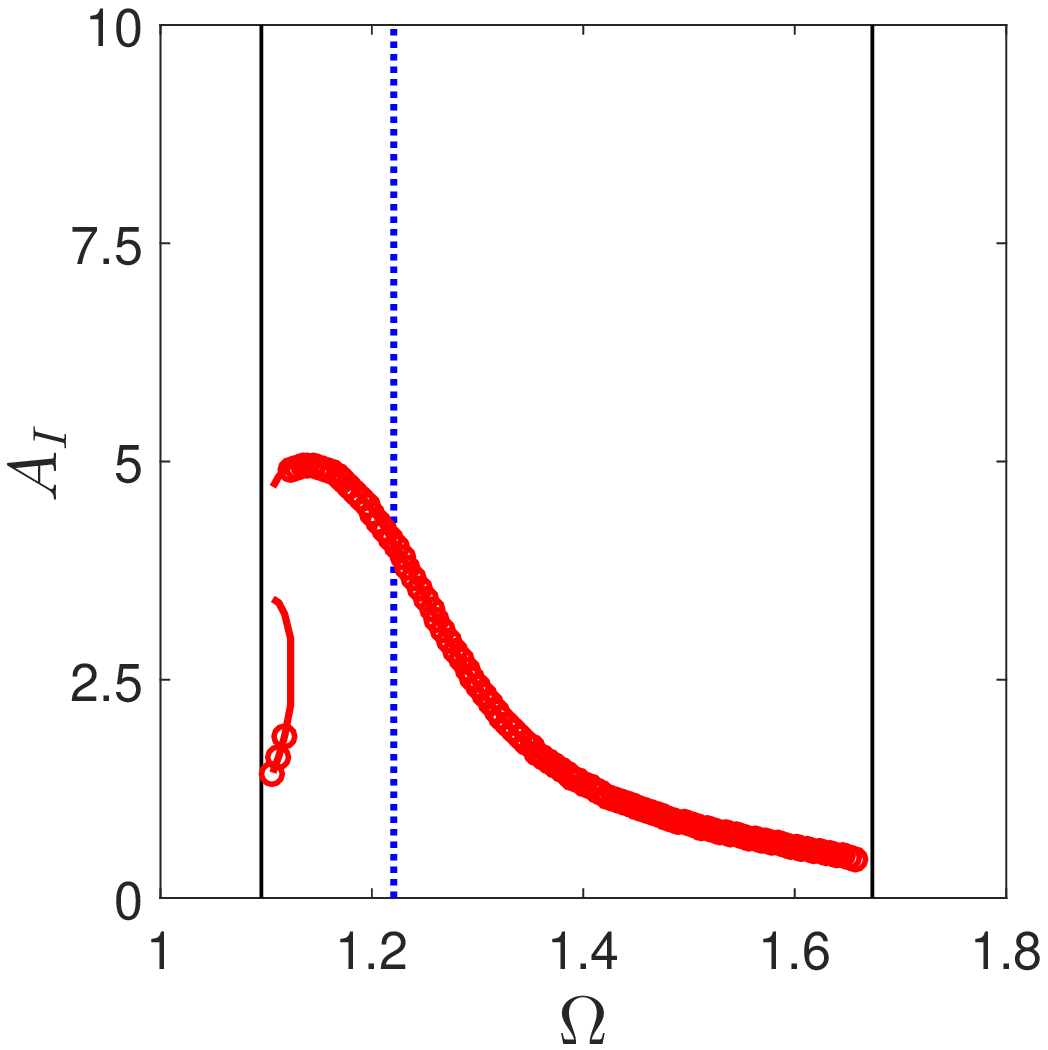}}
	\subfigure[]{
		\includegraphics[width=0.32\textwidth]{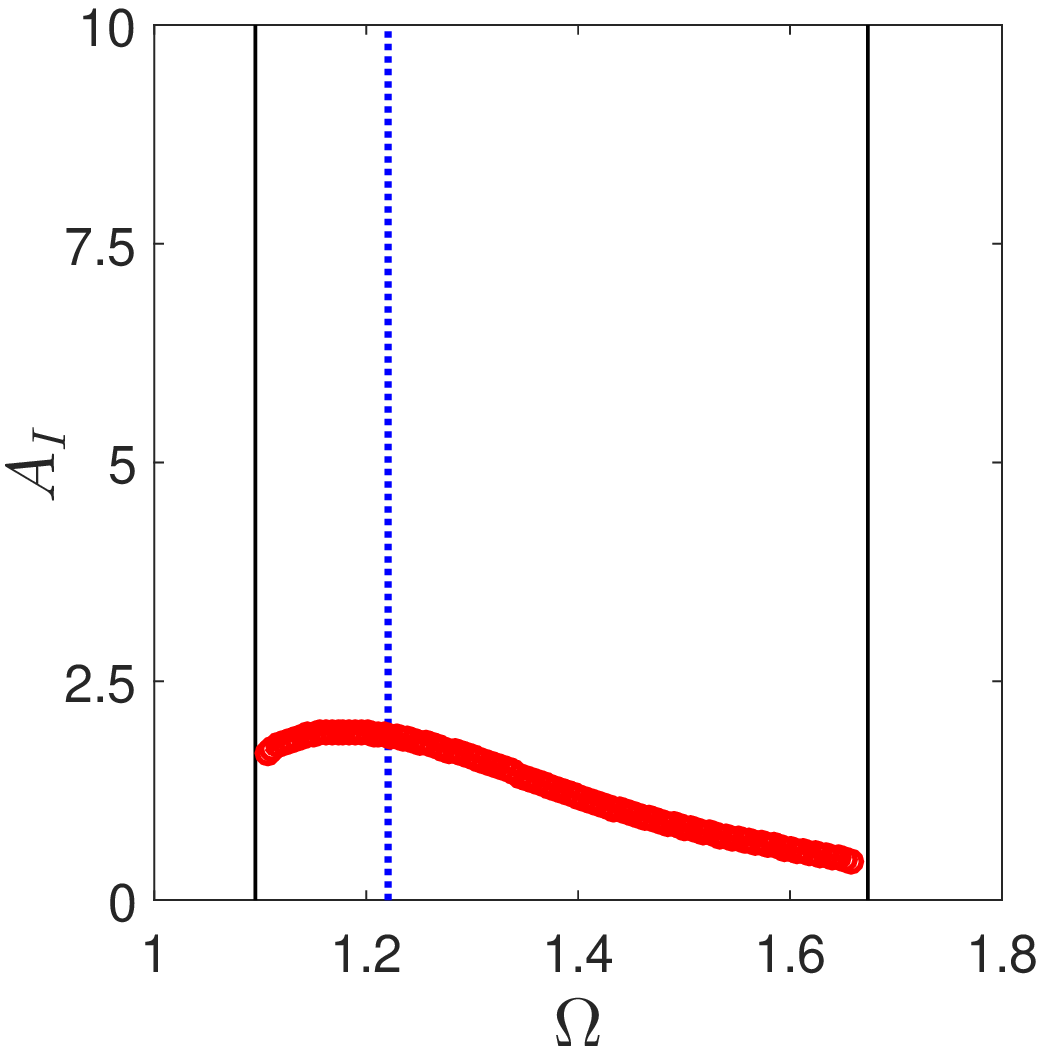}}
	\caption{Analytical (curve) and numerical (circles) responses of a nonlinear chain exited at the interface with force amplitude $f$. 
	The responses for various excitation amplitudes (a) $f=0.001$,  (b) $f=0.06$ and (c) $f=0.2$, are normalized by $f$ and are distinct
	due to nonlinearity.}
	\label{Fig.NLADEM}
\end{figure*}

Let us now exploit the amplitude dependent behavior to migrate the localized mode into the bulk bands. By varying the 
amplitude, the localized mode can be eliminated from the bandgap frequencies. 
The damping coefficient, linear and nonlinear stiffness parameter values are set to 
$\delta = 0.01$, $\gamma = -0.4$ and $\Gamma = -1$, respectively. Notice that strain softening springs ($\Gamma<0$)
are used for this purpose. The 
interface mass in the chain is subjected to the same excitation as in the previous strain 
hardening case. 
Figure~\ref{Fig.NLADEM} displays the frequency response function for the displacement of the interface mass 
predicted by Eqn.~\eqref{DuffingFinal} for three levels of forcing amplitude, (a)~$f = 0.006$, (b)~$f = 0.06$ and (c)~$f = 0.2$. 
The solid vertical lines depict the frequency bounds of the bandgaps, while the dashed (blue) vertical line shows
the frequency of the interface mode when the chain is linear ($\Gamma=0$). 
The markers denote numerical solution obtained by solving the transient problem 
of an equivalent single degree of freedom Duffing oscillator until steady state (with stiffness parameters $k_{e}$ and $\Gamma_{e}$), 
while the solid curves denote the frequency amplitude response of Eqn.~\eqref{DuffingFinal}. 
The interface mode frequency and the normalized amplitude both decrease with increasing force amplitude, 
which is consistent with the behavior of a Duffing oscillator. 
As the amplitude increases, the frequency associated with the interface mode moves into the bulk bands from the bandgaps. 

\begin{figure}[hbtp]
	\centering
	\subfigure[]
	{\includegraphics[width=0.32\textwidth]{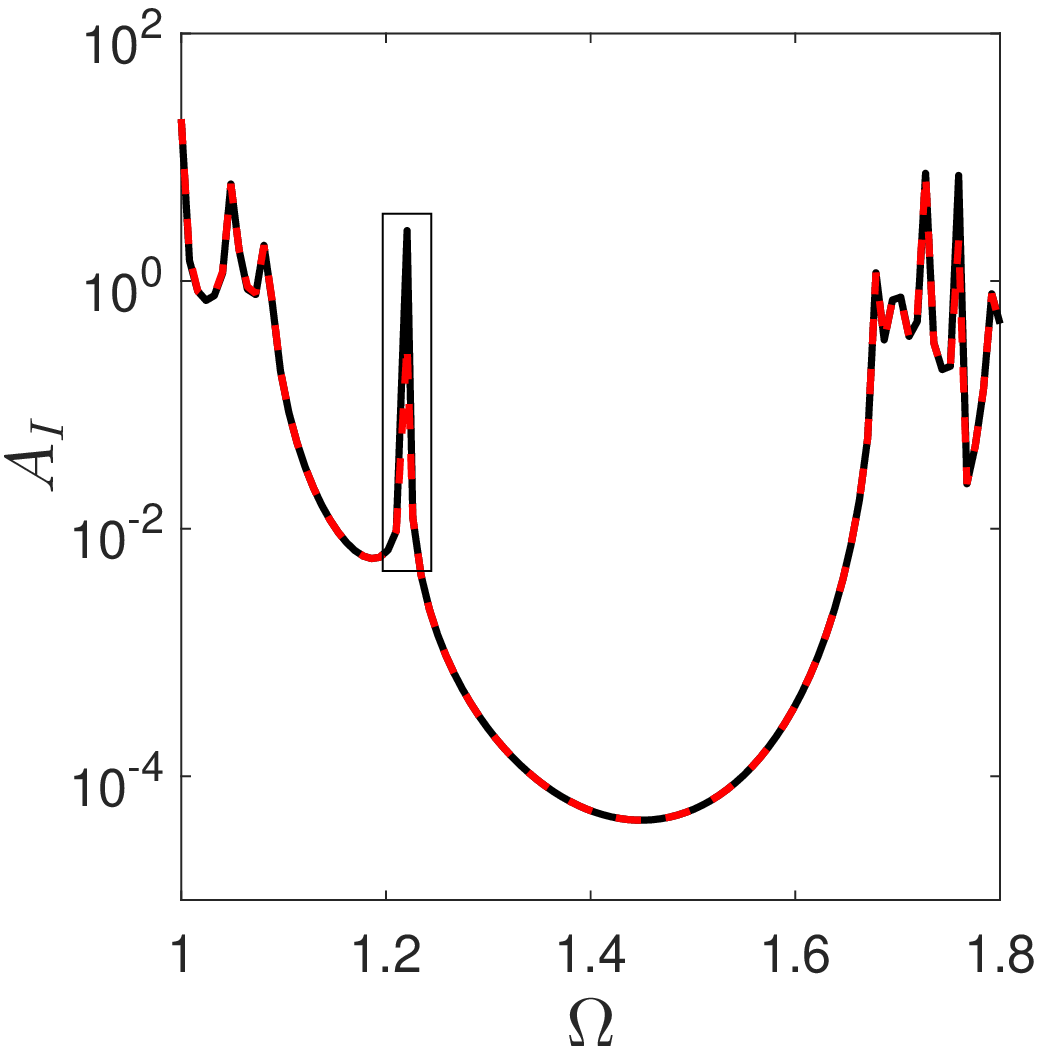}}
	\subfigure[]
	{\includegraphics[width=0.32\textwidth]{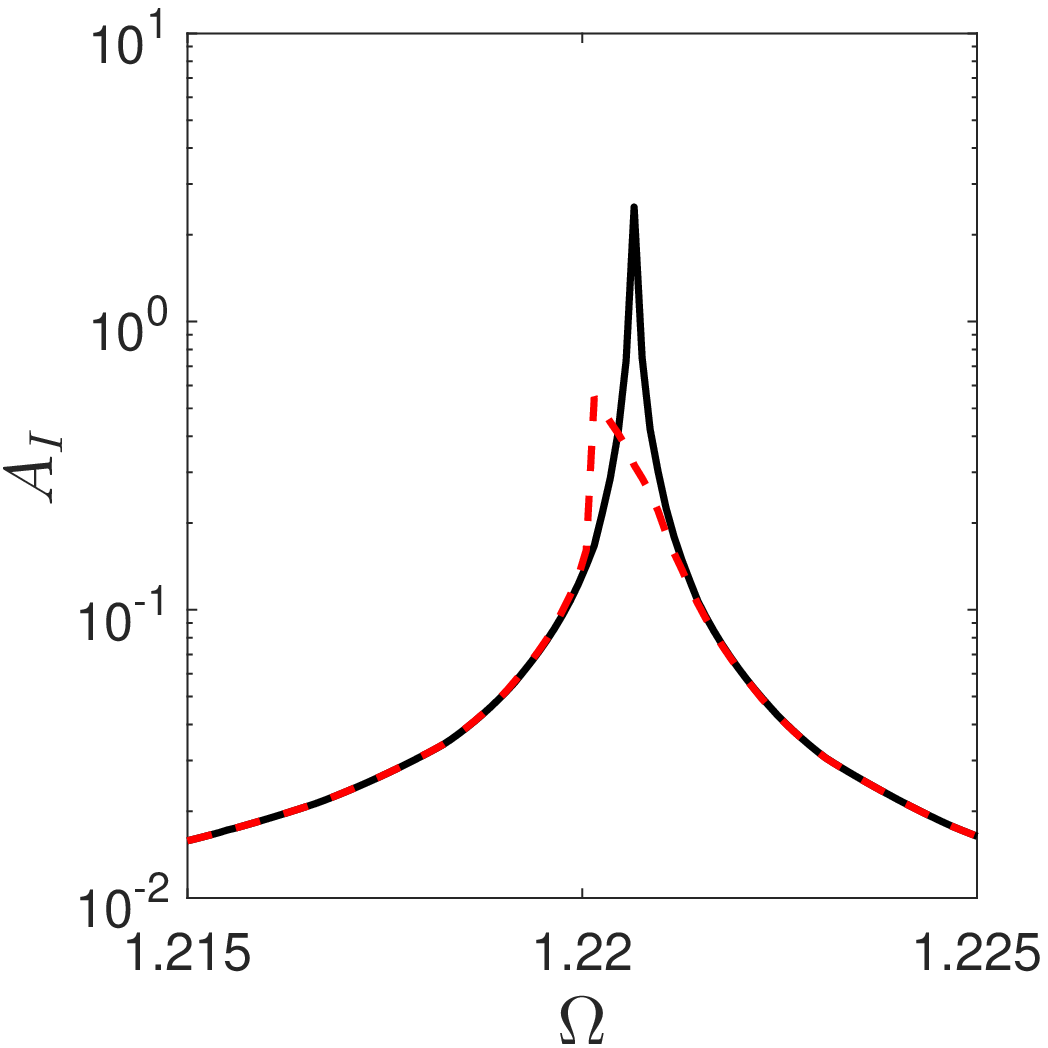}\label{Fig.8b}}
	\caption{Transient response of a nonlinear chain with $\Gamma < 0$ excited at one end. The normalized amplitude response in the bandgaps at high excitation amplitudes (dashed red line, $f=0.06$) is an order of magnitude lower than at low
	amplitudes (solid black line, $f=0.001$).}
	\label{Fig.TunableEM}
\end{figure}

Having demonstrated how to shift the localized mode frequency into the bulk bands using a reduced single degree of freedom model, let us finally show how this shifting leads to a reduction in the response of a finite chain. 
We consider a chain of $20$ unit cells with  an interface mass at the center and subject 
the mass at the left end to a harmonic displacement, while the mass at the right end is free. 
Figure~\ref{Fig.TunableEM} displays the normalized frequency response in the bandgap frequencies 
for two values of excitation force amplitude: $f = 0.001$ (solid curves)
and $f=0.06$ (dashed curves). Figure~\ref{Fig.8b} displays a closeup of the frequency response near the 
interface mode frequency $\Omega_i$. 
The frequency response is similar to the linear case and nonlinear effects are negligible for 
small-amplitude excitations ($f \leq 0.001$), while 
moderate amplitudes ($f>0.01$) lead to a reduction in the displacement amplitude by an order of magnitude. 
The interface mode frequency shifts toward the lower end of the band-gap decreasing the response at the interface. 
Thus the frequency shifting behavior is demonstrated by first showing its analogy with a Duffing 
oscillator using our reduced model and then verifying these predictions with numerical simulations on a finite chain. 
\tred{In summary, amplitude dependent behavior and multiple stable solutions are observed for chains with
stiffness parameter $\gamma < 0$. 
This behavior is predicted analytically by showing the equivalence of the edge mode with a Duffing oscillator. 
Furthermore, edge mode frequency is independent of the wave amplitude for  $\gamma>0$. 
This unexpected observation is explained by examining the analytical solution of eigenmodes associated
with this edge mode. }

%%%%%%%%%%%%%%%%%%%%%%%%%%%%%%%%%%%%%%%%%%%%%%%%%%%%%%%%%%%%%%%%%%%%%%%%%%%%%%%%%%%%%%%%%%%%%%%%%%%%
%%%%%%%%%%%%%%%%%%%%%%%%%%%%%%%%%%%%%%%%%%%%%%%%%%%%%%%%%%%%%%%%%%%%%%%%%%%%%%%%%%%%%%%%%%%%%%%%%%%%
%%%%%%%%%%%%%%%%%%%%%%%%%%%%%%%%%%%%%%%%%%%%%%%%%%%%%%%%%%%%%%%%%%%%%%%%%%%%%%%%%%%%%%%%%%%%%%%%%%%%
%%%%%%%%%%%%%%%%%%%%%%%%%%%%%%%%%%%%%%%%%%%%%%%%%%%%%%%%%%%%%%%%%%%%%%%%%%%%%%%%%%%%%%%%%%%%%%%%%%%%

\section{Tunable edge modes in $2D$ lattices}\label{sec:2Dlattice}
We now extend the ideas presented in the previous section to $2D$ lattices. 
We consider the $2D$ lattice 
in Pal~et.~al.~\cite{pal2016helical} which implements a mechanical analogue of the quantum spin Hall effect and supports topologically protected edge 
modes. An amplitude dependent response is obtained by  using weakly nonlinear springs. 
We present 
dispersion analysis of a unit cell and of an extended unit cell computed using an asymptotic analysis. 
\tred{In contrast to the interface mode in the $1D$ lattice, we show the ability of the considered lattice to 
undergo transitions from bulk-to-edge mode-dominated by varying the excitation amplitude both for 
hardening and softening springs.}
Finally, we present numerical simulations
on finite lattices to illustrate the amplitude dependent nature of wave propagation due to nonlinearities. 

\subsection{Lattice configuration}

\begin{figure}
\centering
\subfigure[]{
\includegraphics[scale=0.2]{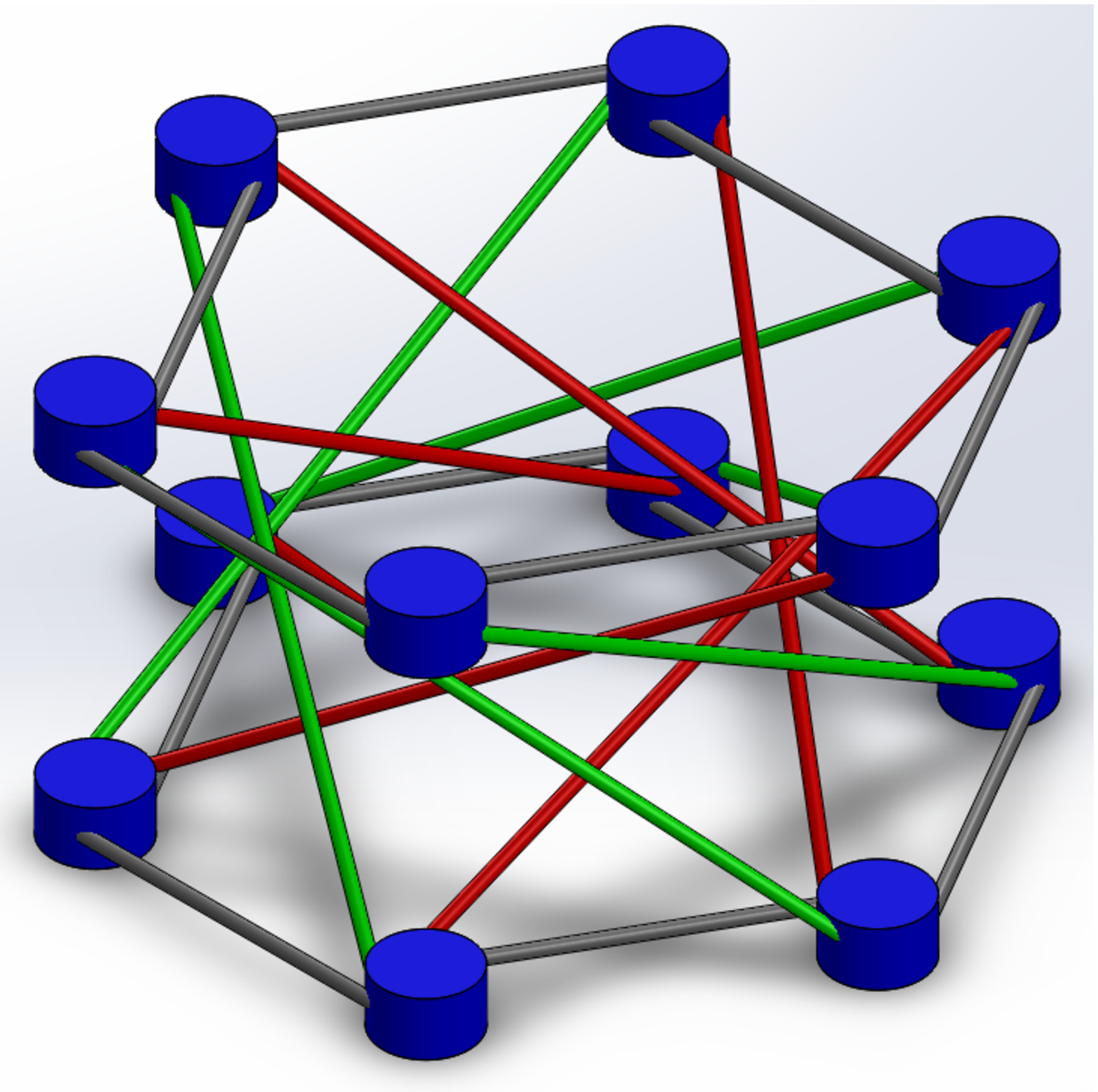}
\label{HexagonalCell}
}
\subfigure[]{
\includegraphics[scale=0.4]{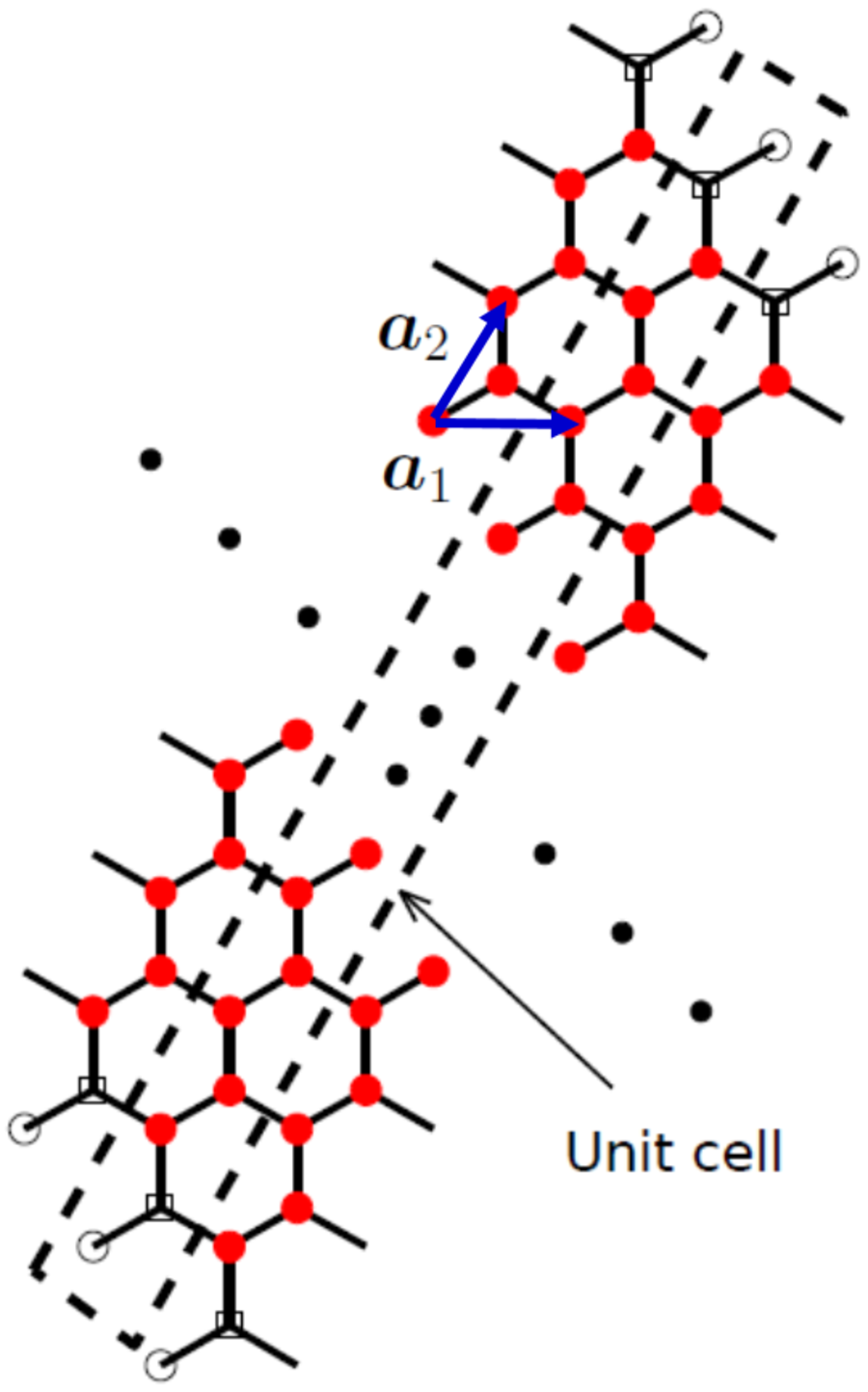}%{Figures/edgeType.eps}
\label{Fig_edgeType}
}
\subfigure[]{
\includegraphics[scale=0.5]{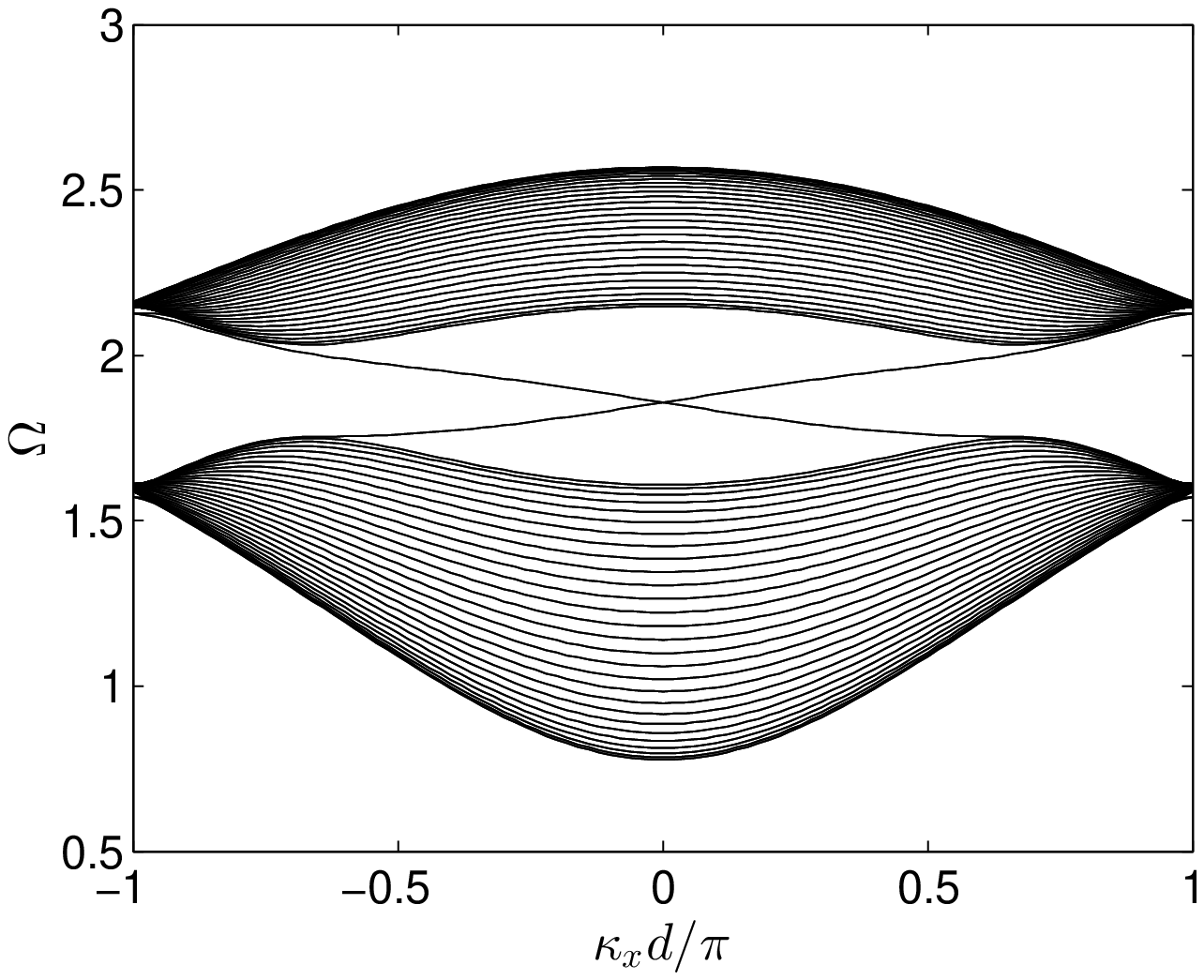}
\label{Linear_EdgeModes}	
}
\caption{(a) A hexagonal cell of the lattice, having two 
layers with normal in-plane springs and a combination of normal and reverse springs between the two layers. 
(b) Finite strip with fixed boundaries. The nodes with filled (red) circles are free, while the others are fixed. 
(c) Dispersion diagram of the finite strip showing edge modes spanning the two sets of bulk modes. 
}
\label{numerStiffSprings}
\end{figure}

The lattice consists of two layers of hexagonal lattice spanning the $xy$-plane. 
and its lattice vectors are $\ba_1 = (-1/2,  \sqrt{3}/2)$ and $\ba_2 = (1/2, \sqrt{3}/2)$. 
Figure~\ref{HexagonalCell} displays a schematic
of a single hexagonal cell. Each node is a disk that rotates about the $z$-axis, perpendicular to the 
plane of the lattice. Two kinds of springs,  
normal and chiral, connect the disks. The in-plane springs (gray color in Fig.~\ref{HexagonalCell}) 
are linear and they provide a torque $k(\theta_j-\theta_i)$
on disk $i$ due to rotations $\theta_i$ and $\theta_j$ of the two nearest neighbor disks connected to the spring. 
A combination of normal ($n$, green color in Fig.~\ref{HexagonalCell}) 
and chiral ($ch$, red color in Fig.~\ref{HexagonalCell}) springs connect the second nearest neighbors on adjacent layers in our lattice. 
These springs are weakly nonlinear and the 
torque-rotation relation between two disks $i,j$ are, respectively,
\begin{subequations}
\begin{align}
T^n_i &= k_n (\theta_j - \theta_i) + \epsilon_n (\theta_j - \theta_i)^3 , \\ 
T^{ch}_i &= -k_{ch} (\theta_j + \theta_i) - \epsilon_{ch} (\theta_j + \theta_i)^3 . 
\end{align}
\end{subequations}

\subsection{Dispersion analysis of linear and nonlinear lattices}\label{Sec.Disp2D}

Dispersion studies are conducted 
both for a single hexagonal unit cell having $4$ degrees of freedom ($2$ in each layer) and for a unit cell of  
a strip which is periodic along one direction, as illustrated in Fig.~\ref{Fig_edgeType}. 
Let us set $\bu$ as the vector whose components are the generalized 
displacement for all the degrees of freedom in a unit cell, which in our case, 
would be the rotation of disks at each lattice site. 
In~\cite{pal2016helical}, the authors show that this lattice has a band gap for bulk modes. Furthermore, there are topologically 
protected edge modes in this band gap which propagate along the boundaries of the lattice. We seek to investigate how weak nonlinearities 
affect the edge modes in our lattice.  
To get the dispersion relation of a nonlinear lattice, we use a perturbation based method to seek corrections to the linear dispersion
relation $\omega = \omega(\bmu)$, with $\bmu$ being the two dimensional wavevector. 
Based on the method of multiple scales, the following asymptotic expansion for the displacement components in a unit cell and frequency
is imposed
\begin{gather*}
\bu = \bu_0 + \epsilon \bu_1 + O(\epsilon^2), \\ 
\omega = \omega_0 + \epsilon \omega_1 + O( \epsilon^2). 
\end{gather*}
The asymptotic procedure we follow is similar to Leamy and coworkers~\cite{narisetti2010perturbation,Narisetti2011} 
and its details are presented in Appendix~\ref{asymptoticTheory}. 

We first present the dispersion behavior of a finite strip 
of a linear lattice to illustrate the existence of localized edge modes. Then, two kinds of nonlinear springs, strain 
hardening and strain softening, are considered to demonstrate the amplitude dependent nature of these edge modes. 
The equations are normalized using the time scale $\sqrt{k/I}$, with $I$ being the rotational inertia of 
the disks. In non-dimensional form (with superscript $\tilde{k}$), both the normal and chiral springs connecting adjacent layers
are chosen to have a linear stiffness component $\tilde{k}_n = \tilde{k}_{ch} = 0.1$ and their
nonlinear components are equal ($\epsilon = \tilde{\epsilon}_n = \tilde{\epsilon}_{ch}$).

\subsubsection{Dispersion analysis of a strip}

To illustrate the presence of edge modes in our lattice, let us consider a finite strip of $20$ unit cells as illustrated in 
Fig.~\ref{Fig_edgeType}. The strip is periodic in the $\ba_1$ direction and has a finite width in the $\ba_2$ direction. 
The nodes with red (filled circle) markers are free to move, while the nodes with unfilled circles and squares at either boundary are fixed 
nodes. A dispersion analysis is conducted on this finite strip which is periodic in the $\ba_1$ direction and the dashed 
rectangle shows the unit cell. By imposing a traveling wave solution of the form $\bu = \bu(\kappa_x) e^{i(\Omega t - \kappa_x\cdot x)}$
on the lattice, an eigenvalue problem is obtained for each
wavenumber $\kappa_x$. Note that the $x$-axis is  oriented along the $\ba_1$ direction. 
 
Figure~\ref{Linear_EdgeModes} displays the dispersion diagram for the finite strip under study. The wavenumber $\kappa_1$ is projected onto
the $x$-axis. There are two sets of wave modes: the first set spans $[0.78,\; 1.75]$ and the second set spans $[2.03 , \; 2.55]$. These two sets 
correspond to bulk modes and the two modes between them are edge modes. The eigenvectors corresponding to these frequencies are localized 
at the edges. 
We remark here on the choice of boundary conditions as shown in Fig.~\ref{Fig_edgeType}. Note that allowing the nodes with square markers 
to be free results in a different type of edge mode than the one illustrated in Fig.~\ref{Fig_edgeType}. The work in~\cite{pal2016helical} presented the band diagrams when the nodes having square markers were not fixed.  There are two overlapping bands at each point in the dispersion diagrams in Fig.~\ref{Linear_EdgeModes}. The lattice supports two traveling waves at the edge of the lattice: one in 
the clockwise and the other in the counter-clockwise direction. Furthermore, these modes are topologically protected: they span the 
entire bandgaps and they cannot be localized by small disorders or perturbations~\cite{hasan2010colloquium}.

\subsubsection{Strain hardening springs}\label{Sec.Hardening}

\begin{figure}
\centering
\subfigure[]{
\includegraphics[scale=0.6]{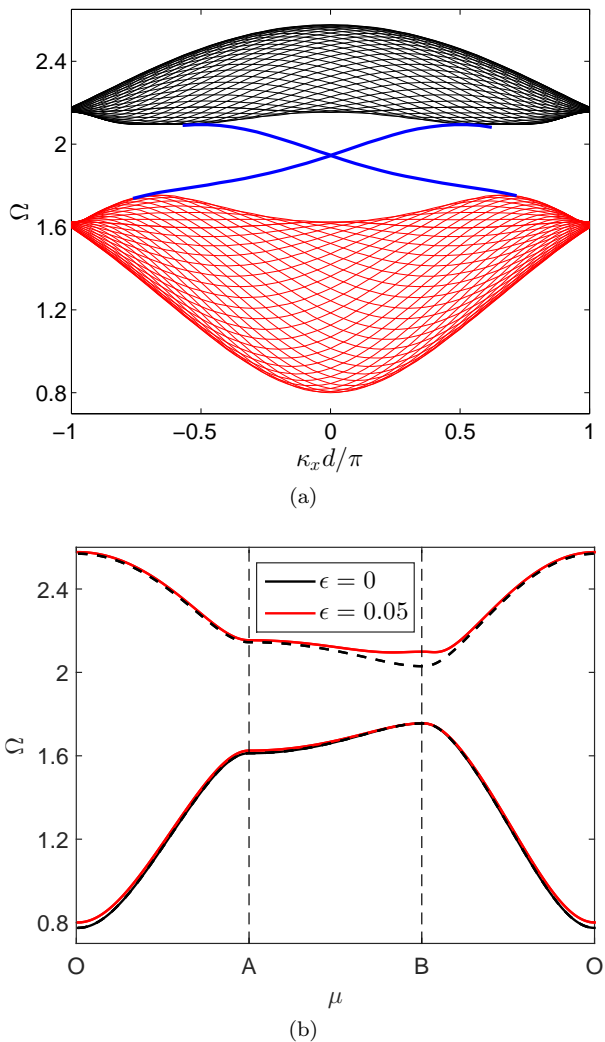}
\label{stiff_stripEdgeDisp}
}\vspace{0.12in}
\subfigure[]{
\includegraphics[scale=0.6]{IBZ_stiff_A6e-1.eps}
\label{stiff_IBZ}
}
\caption{Dispersion diagram for both linear ($\epsilon=0$) and strain hardening springs $\epsilon=0.05$ over 
(a) a strip and (b) the irreducible Brillouin zone.  
The edge modes traverse the bandgaps and the optical band shifts upward near the point $B$ in the nonlinear lattice.  
}
\label{stiff_Disp}
\end{figure}

Having demonstrated the presence of edge modes in a linear lattice, we now investigate the effect of introducing nonlinear interactions between
the interlayer springs. Figure~\ref{stiff_Disp} displays the dispersion diagram when the nonlinearity is of the 
strain hardening type $(\epsilon = 0.05)$ with amplitude of the waves $A_0 =0.6$. 
The first order correction is computed using an asymptotic analysis 
(Eqn.~\eqref{Eqn.Omeg1} in Appendix~\ref{asymptoticTheory}) at each two dimensional wavevector $\bmu$ for both a unit
cell in the bulk and a unit cell comprised of a finite strip. Figure~\ref{stiff_stripEdgeDisp} displays the bulk 
dispersion surface projected onto the $x$-axis along with the edge modes computed from the finite strip. A comparison with the dispersion
diagram of the finite strip in the linear case shows that the lower band remains unchanged while the lower surface of the upper 
band shifts upward. 

Figure~\ref{stiff_IBZ} displays the dispersion curves along the boundary of the irreducible Brillouin zone for both 
the linear (dashed curves) and nonlinear (solid curves) lattices. Since the hexagonal lattice has a six-fold symmetry, the IBZ 
is a triangle and we choose it to span the points  $O:(0,0)$, $A: (0,\pi)$ and $B:(2\pi/3,2\pi/3)$ in the reciprocal lattice space. 
The presence of inter-planar springs leads to 
a bandgap for bulk waves, as shown in Fig.~\ref{stiff_IBZ} and the existence of edge waves in this bandgap.
We see that the lower band does not get significantly affected due to the nonlinear springs. However, the upper band in the vicinity 
of point $B$ gets shifted upward and the bandgap widens as a consequence. Note that the edge modes continue to span the bandgaps 
and they do not localize (group velocity is nonzero) in the presence of nonlinear interactions. 

We now elaborate how the above observations can be exploited to achieve amplitude dependent edge waves using our nonlinear lattices. 
At small amplitudes, the dynamic response is similar to a lattice with no nonlinear springs and corresponds to the $\epsilon = 0$ case 
in Fig.~\ref{stiff_Disp}. 
However as  the amplitude increases, nonlinear effects come into play and the behavior resembles 
the nonlinear case, illustrated by $\epsilon = 0.05$ in Fig.~\ref{stiff_Disp}. 
Thus exciting at a frequency at the tip of the lower surface of the Brillouin zone near point $B$ will result in 
amplitude dependent edge waves. At small amplitudes, there will be no  edge waves, while at high amplitudes, the band widens and one-way
edge waves propagate in the lattice. 

\subsubsection{Strain softening springs}\label{Sec.Softening}

\begin{figure}
\centering
\subfigure[]{
\includegraphics[scale=0.6]{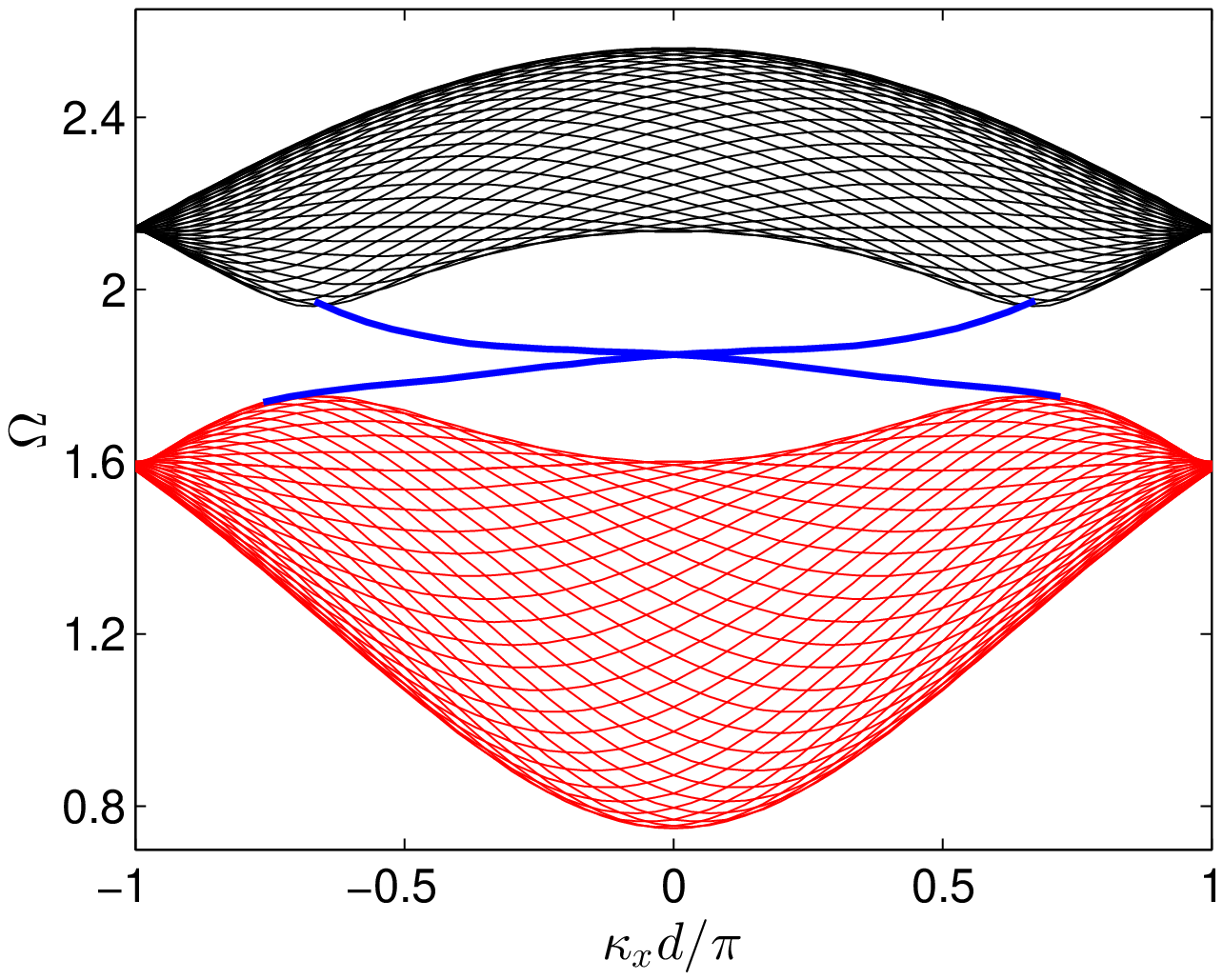}
\label{soft_stripEdgeDisp}
}
\subfigure[]{
\includegraphics[scale=0.6]{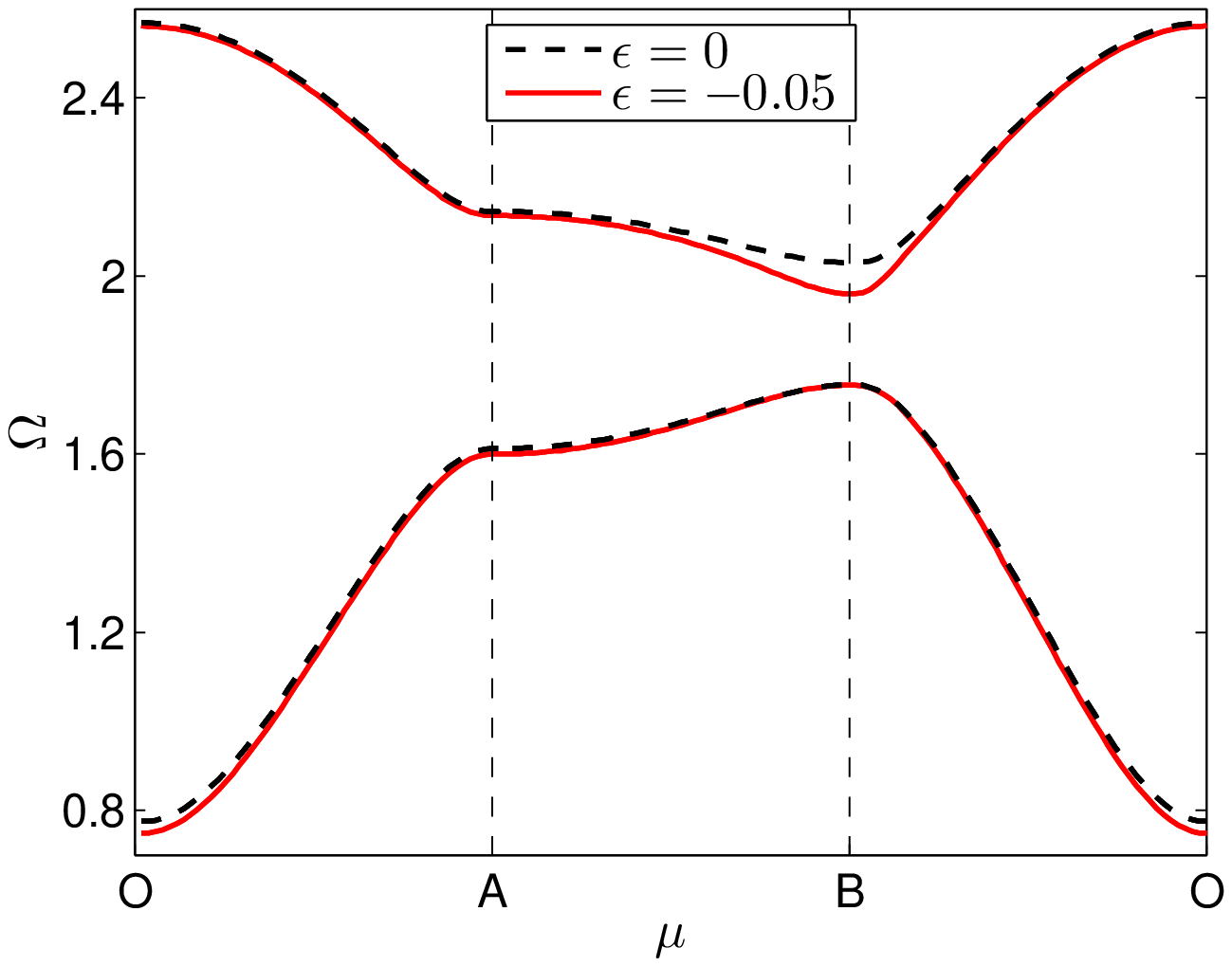}
\label{soft_IBZ}
}
\caption{Dispersion diagram for lattices with linear ($\epsilon=0$) and strain softening springs $\epsilon=-0.05$ over 
(a) a strip and (b) the irreducible 
Brillouin zone. The edge modes traverse the bandgaps and the optical band shifts downward near the point $B$ in the nonlinear lattice. 
}
\label{soft_Disp}
\end{figure}

We now turn attention to the study of nonlinear springs of the strain softening type having $\epsilon < 0$. 
A similar dispersion analysis is conducted on both a strip and a single unit cell with nonlinear stiffness parameter $\epsilon=-0.5$ 
and  wave amplitude $A_0 = 0.6$. 
Figure~\ref{soft_stripEdgeDisp} displays the dispersion surface of both the bulk and edge modes projected onto the $x$-axis. Similar to the 
earlier case with $\epsilon > 0$, the lower band does not change significantly due to the nonlinear springs. The lower surface of the upper 
band shifts downward, which is consistent with the behavior for $\epsilon > 0$ case, since the first order correction 
$\epsilon\omega_1$ is linearly proportional to $\epsilon$. Figure~\ref{soft_IBZ} displays the dispersion curves along the 
boundary of the IBZ. In contrast with the strain hardening case, here the dispersion curves shift downward near the point $B$ while 
remaining relatively unaltered away from this point. 
Similar to the strain hardening case, these softening springs can be exploited to get amplitude dependent response of the lattice. The 
lattice behavior can be changed from edge waves at low amplitudes to bulk waves at high amplitudes. 
We thus illustrated the amplitude dependent nature of the dispersion curves for the strain softening nonlinear springs.

\subsection{Numerical simulations of wave propagation}
We now conduct numerical simulations to demonstrate the effect of nonlinear interactions on wave propagation in a finite 
lattice. The numerical results are interpreted using the dispersion diagrams for linear and nonlinear lattices 
that were presented earlier 
in Sec.~\ref{Sec.Disp2D}. All our numerical simulations are conducted on a lattice of $30\times 30$ unit cells using 
a fourth order Runge Kutta explicit time integration scheme. The boundary nodes of our lattice are fixed similar to that
illustrated in Fig.~\ref{Fig_edgeType}. 
The lattice is subjected to a point excitation at a specific frequency 
on a boundary node lying at the center of the lower left boundary. 
Two examples are presented: the first one demonstrates edge wave propagation
at high amplitudes, while the second example demonstrates the decaying of edge waves with increasing amplitude.

\subsubsection{High amplitude edge waves}\label{sec:2Dlattice_numer}

\begin{figure}
\centering
\subfigure[$\;A = 6\times10^{-3}, \;\tau=200$]{
\includegraphics[scale=0.4]
%{Figures/FigTrialE_4000.eps}
{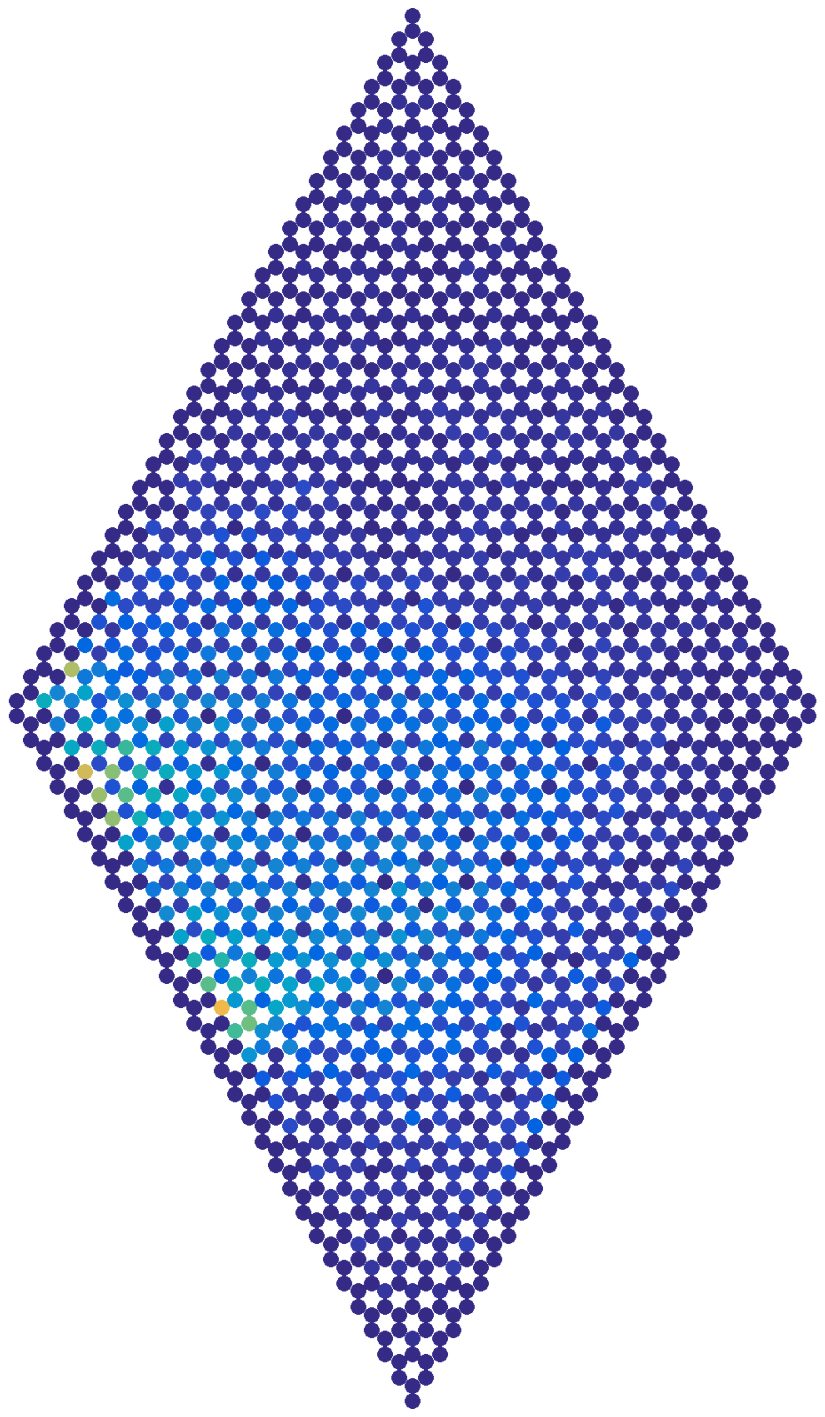}
\label{stiffSpring_Low1}
}
\subfigure[$\;A = 6\times10^{-3}, \;\tau=400$]{
\includegraphics[scale=0.4]
%{FigTrialE_8000.eps}
{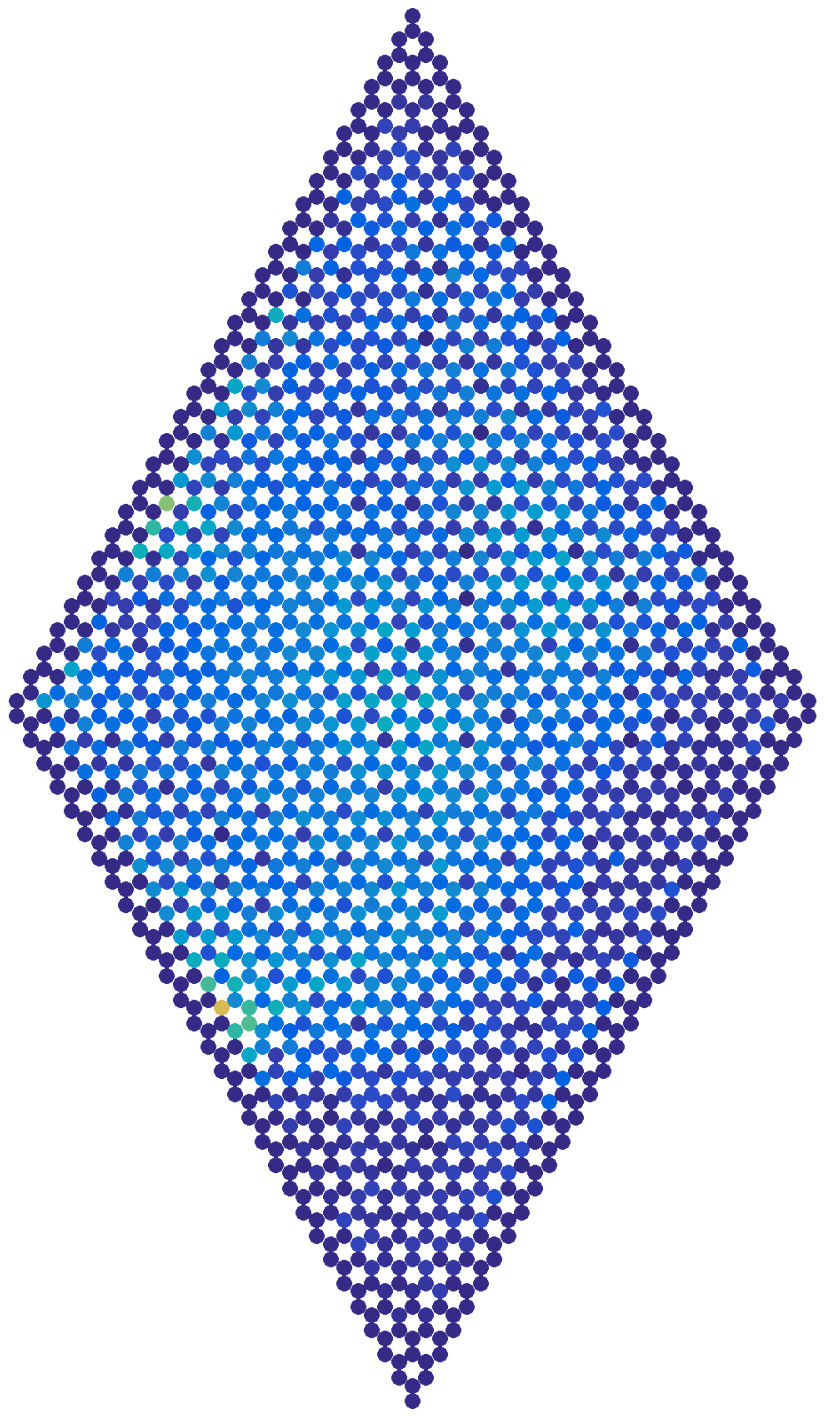}
\label{stiffSpring_Low2}
}\\
\subfigure[$\;A = 6\times10^{-1}, \;\tau=200$]{
\includegraphics[scale=0.4]
%{FigTrialD_4000.eps}
{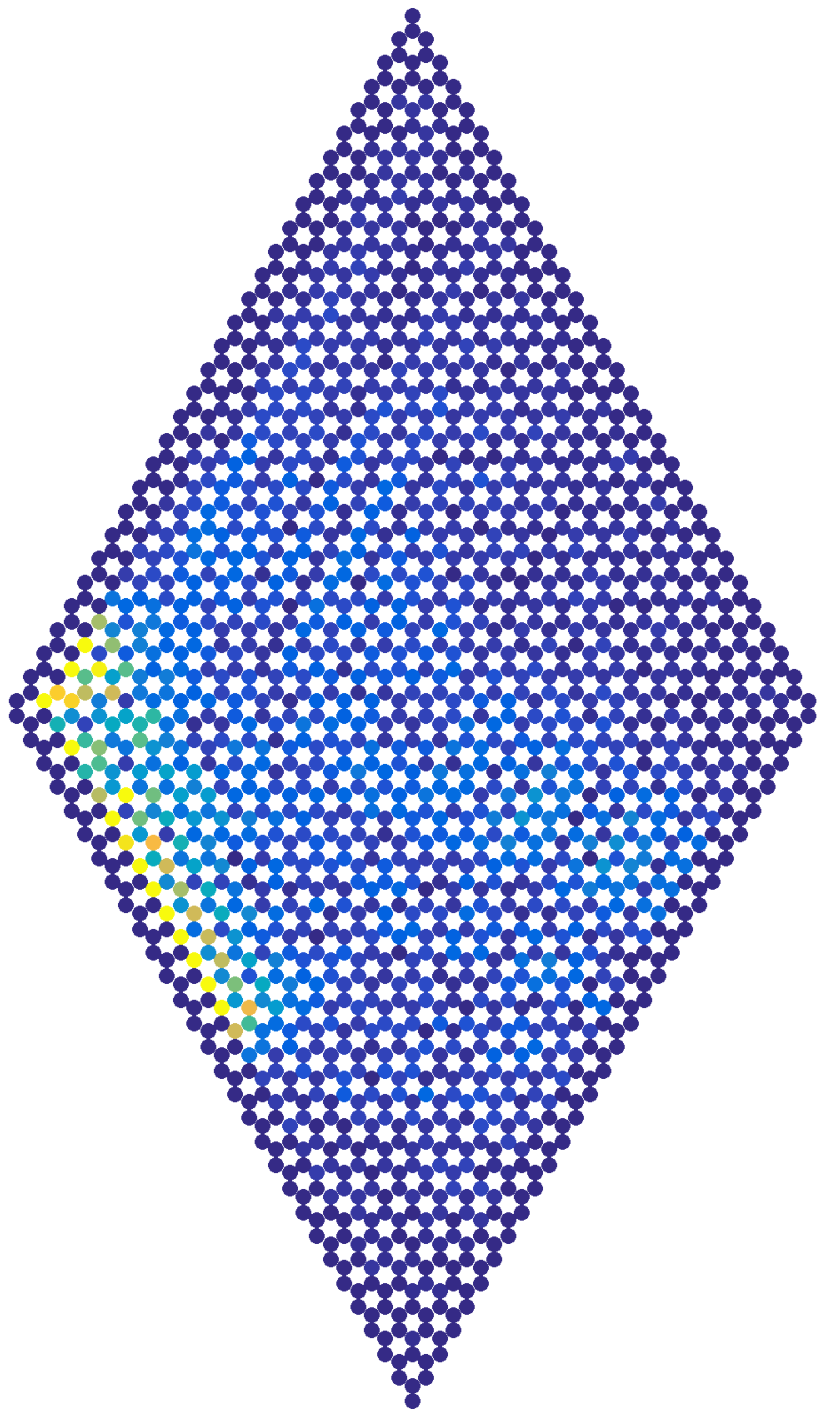}
\label{stiffSpring_High1}
}
\subfigure[$\;A = 6\times10^{-1}, \;\tau=400$]{
\includegraphics[scale=0.4]
%{FigTrialD_8000.eps}
{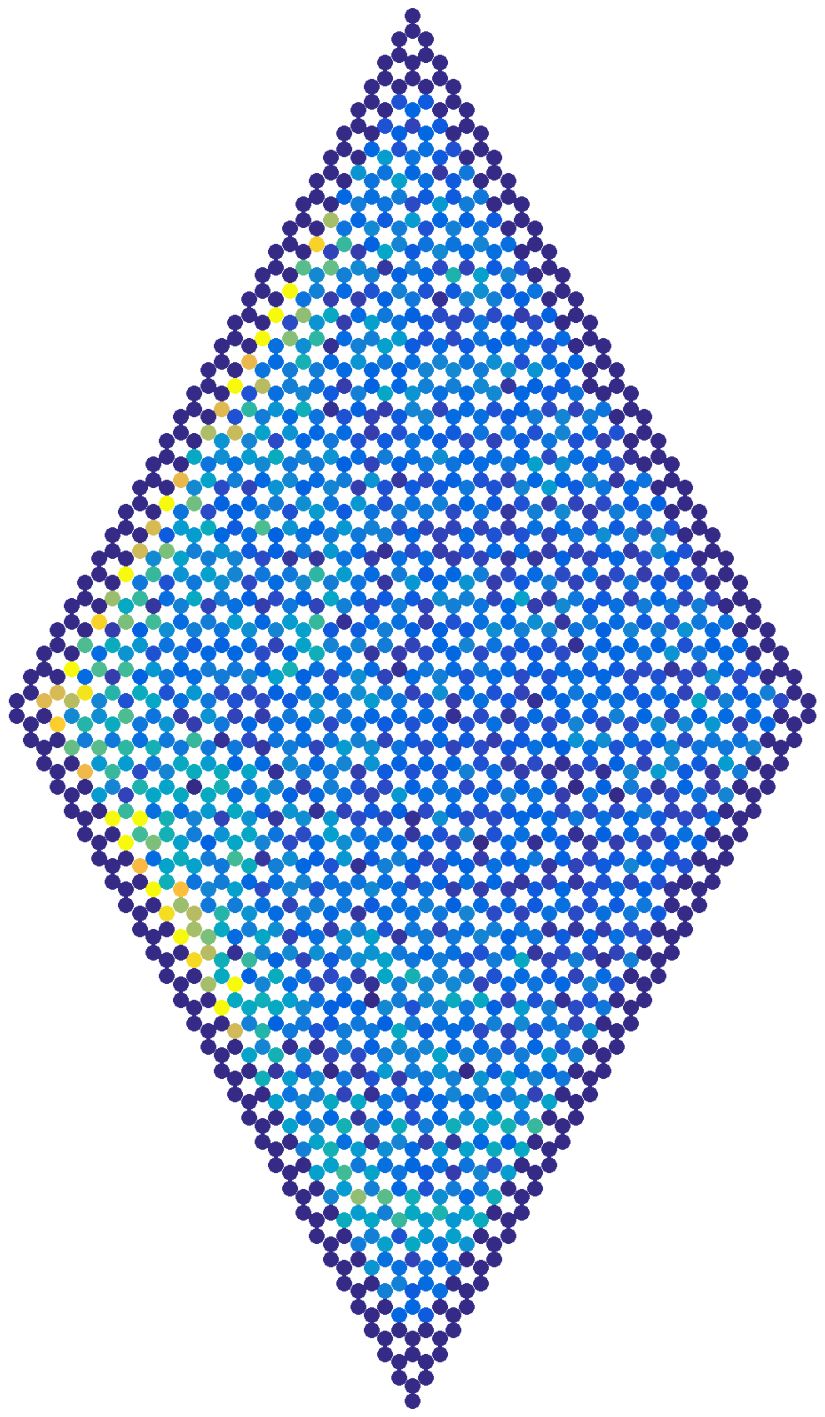}
\label{stiffSpring_High2}
}
\caption{Displacement magnitude at each lattice site at two time instants for lattice with $\epsilon = 0.05$ subjected to two force amplitudes (a,b) $A = 6\times10^{-3}$, (c,d) $A = 6 \times 10^{-1}$ and excitation frequency $\Omega= 2.045$. The lattice supports bulk and edge waves at 
low and high amplitudes, respectively. 
}
\label{numerHardSprings}
\end{figure}
%\begin{figure}
%\centering
%\subfigure[]{
%\includegraphics[scale=0.6]{Figures/FigFile_D}
%\label{stiffSpring_Low}
%}
%\subfigure[]{
%\includegraphics[scale=0.6]{Figures/FigFile_E}
%\label{stiffSpring_High}
%}
%\caption{Displacement at $\tau= 200$ for lattice with $\epsilon = 0.05$ subjected to force amplitudes (a) $A = 6\times10^{-3}$ and 
%(b) $A = 6 \times 10^{-1}$ and excitation frequency $\Omega= 2.045$. The lattice supports bulk and edge waves at 
%low and high amplitudes, respectively. 
%}
%\label{numerHardSprings}
%\end{figure}

In this example, the lattice comprises of nonlinear springs of the strain hardening type with $\epsilon = 0.05$. 
Two numerical simulations are conducted: one at low ($A = 6\times 10^{-3}$) and the other at high ($A = 6\times 10^{-1}$)
force excitation amplitudes. A boundary lattice site is subjected to 
to a harmonic excitation at frequency $\Omega=2.045$. 
The dispersion analysis in Fig.~\ref{stiff_Disp} shows that this 
frequency lies in the lower part of the top band and  
the linear lattice supports bulk waves and no edge waves. 
As discussed earlier in Sec.~\ref{Sec.Hardening}, at higher amplitudes, the bandgap widens and edge modes exist at higher frequencies. 
The top and bottom layers are subject to the excitation 
\begin{equation}\label{EqLoading}
F_{top} = A \cos \omega t,  \;\; F_{bottom} = A \sin \omega t. 
\end{equation}

Figure~\ref{numerHardSprings} displays the angular displacement of the disks at the various nodes. The color scale ranges for the two cases
are $[0,8\times 10^{-3}]$ and $[0,8\times 10^{-1}]$. The colors show the 
magnitude ($l_2$ norm) of the displacement vector at each lattice site. 
Note that there are two disks (one at top and  one at bottom layer)  
at each lattice site and the displacement vector  thus has two components 
denoting the angular displacement of these two disks. Figures~\ref{stiffSpring_Low1}-(b) 
display the  displacement magnitude 
for the low amplitude excitation at times $\tau = 200$ and $\tau=400$. It is observed that the wave propagation is 
isotropic from the point of excitation into the lattice and this behavior is consistent with the predictions of the dispersion
analysis as there are no edge waves at the excitation frequency. 
Figures~\ref{stiffSpring_High1}-(d) displays the displacement magnitude for high amplitude excitation at the same time instants. 
Edge waves are observed to propagate in the counter-clockwise 
direction, which is indeed consistent with the behavior predicted in the dispersion analysis in Sec.~\ref{Sec.Hardening}. 

\subsubsection{Bulk waves at high amplitudes}

\begin{figure}
\centering
\subfigure[$\;A = 2\times10^{-3}, \tau=400$]{
\includegraphics[scale=0.4]
%{Figures/FigTrial_8000}
{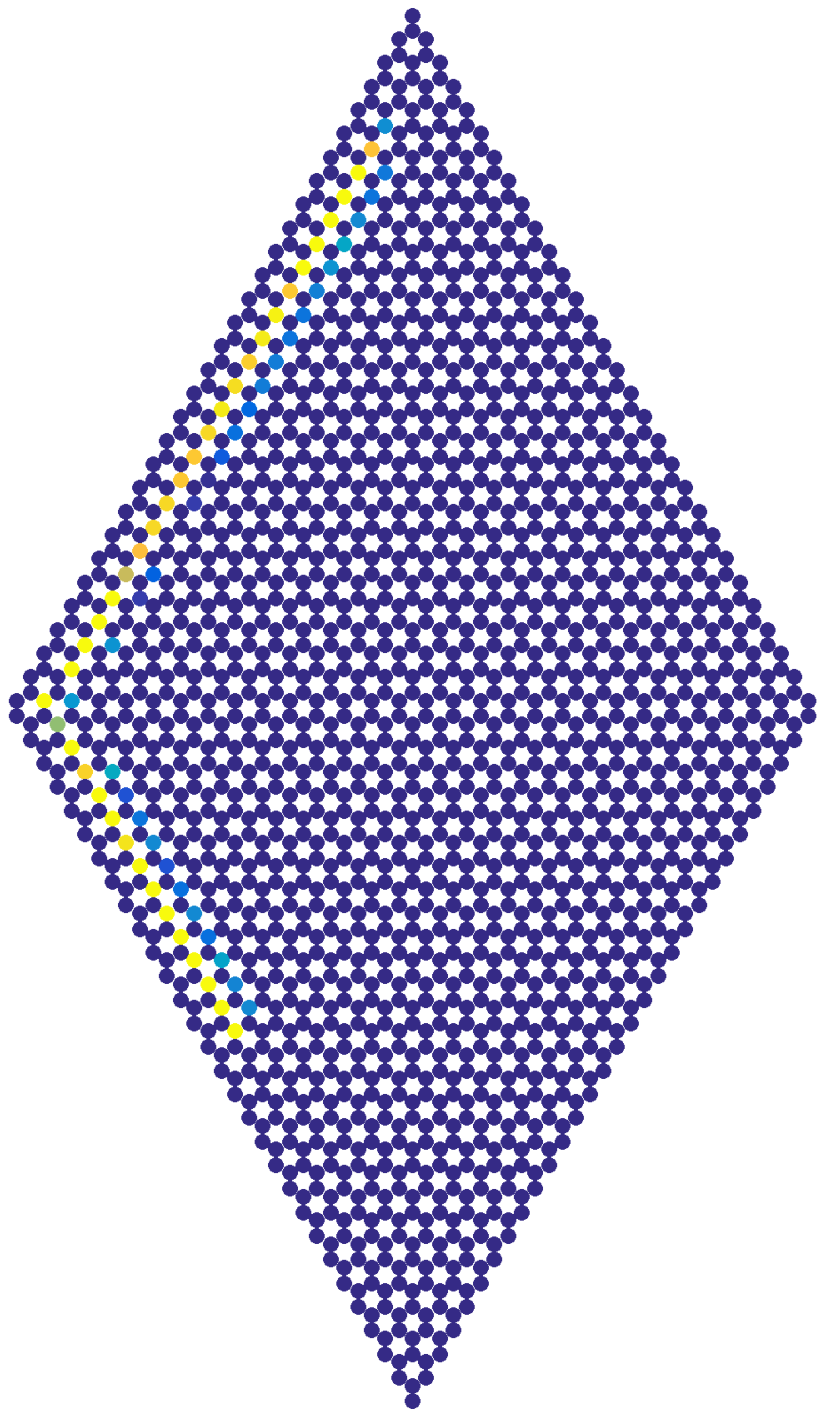}
\label{softSpring_Low1}
}
\subfigure[$A = 2\times10^{-3}, \tau=1000$]{
\includegraphics[scale=0.4]
%{Figures/FigTrial_20000}
{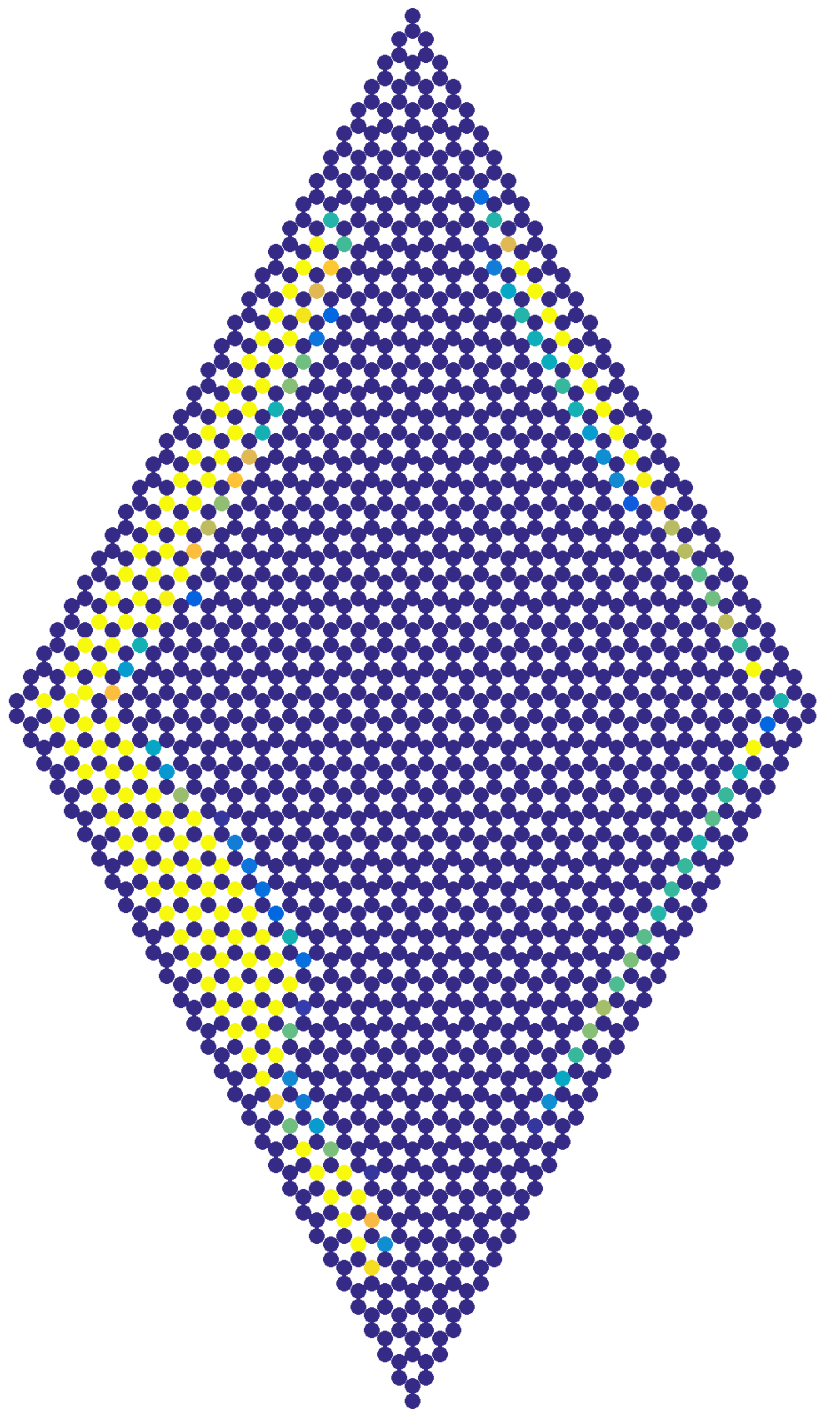}
\label{softSpring_Low2}
}\\
\subfigure[$\;A = 2\times10^{-1}, \tau=400$]{
\includegraphics[scale=0.4]
%{Figures/FigTrialB_8000}
{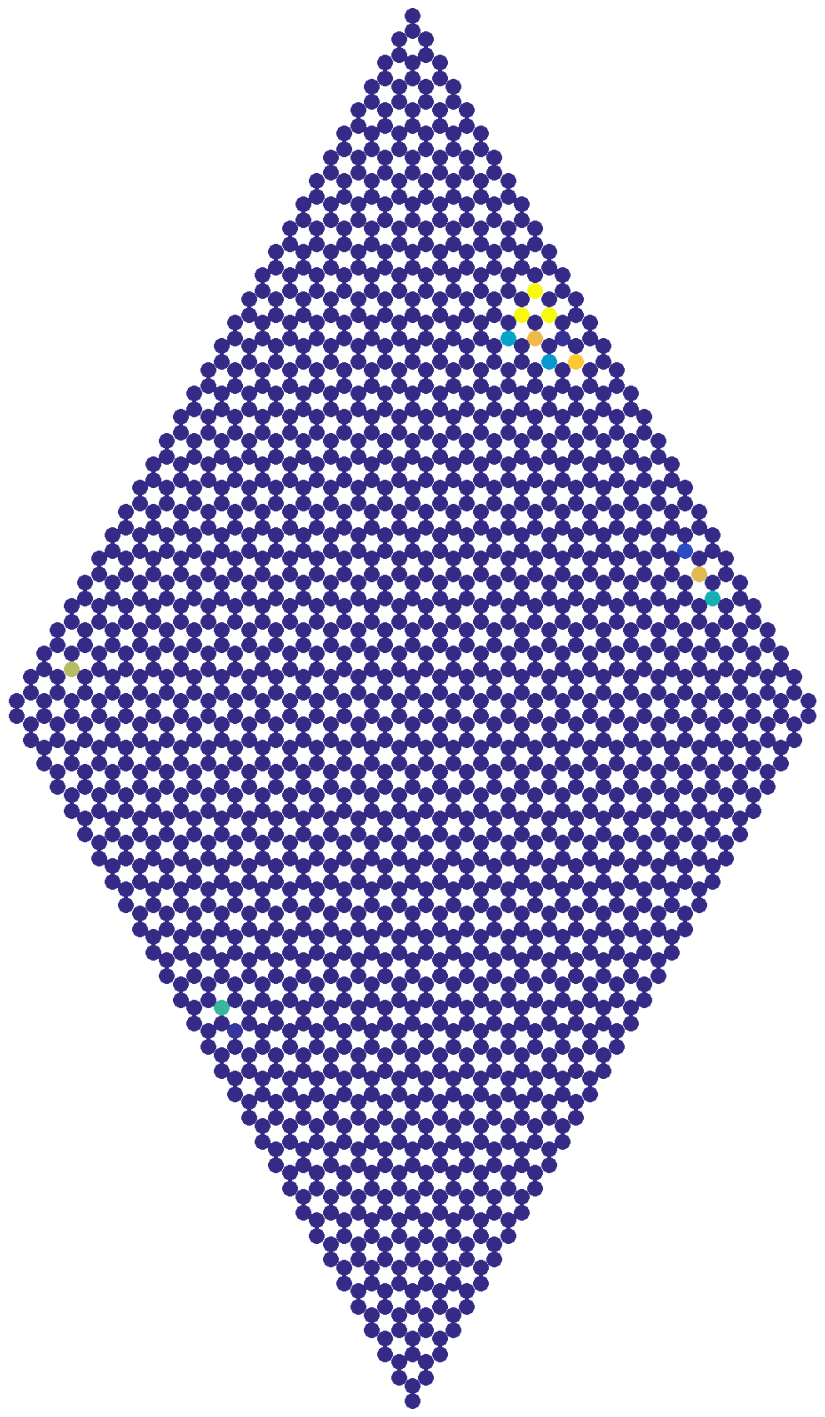}
\label{softSpring_High1}
}
\subfigure[$A = 2\times10^{-1}, \tau=1000$]{
\includegraphics[scale=0.4]%{Figures/FigTrialB_20000}
{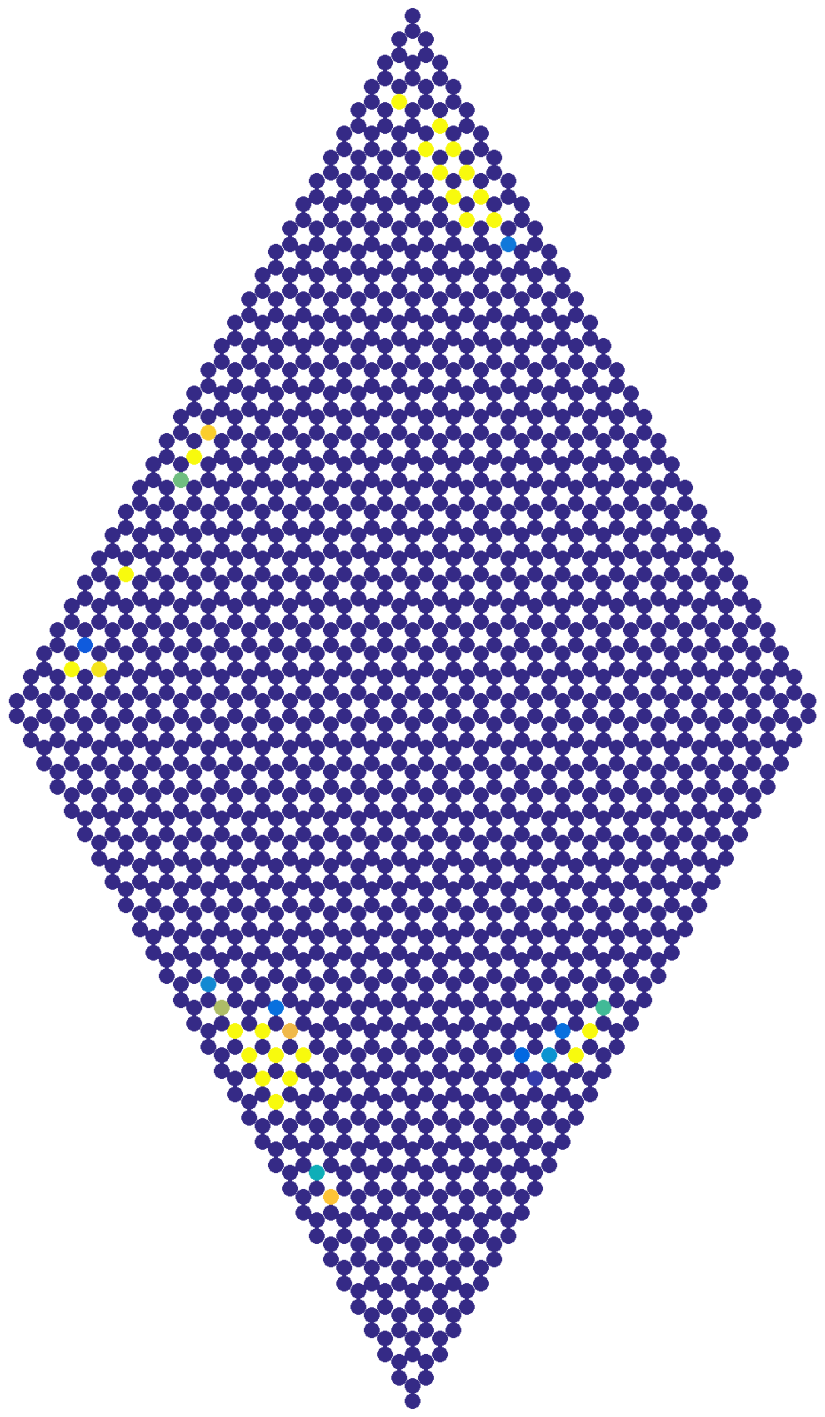}
\label{softSpring_High2}
}
\caption{
Displacement magnitude at each lattice site at two time instants for lattice with $\epsilon = 0.05$ subjected to two force 
amplitudes (a,b) $A = 2\times10^{-3}$, (c,d) $A = 2 \times 10^{-1}$ and excitation frequency $\Omega= 2.02$. 
The lattice supports edge waves at low amplitudes and bulk waves at 
high amplitudes. The color scale has a range (a,b): $[2.4\times 10^{-3}, \;3\times 10^{-3}]$ and (c,d): $[2.4\times 10^{-1}, \;3\times 10^{-1}]$. 
}
\label{numerSoftSprings}
\end{figure}

Our next example involves strain softening springs having $\epsilon < 0$. Again, the lattice is subjected to a point excitation at a 
frequency $\Omega = 2.02$ with the top and bottom disks at the node having a phase difference of $\pi/2$ as in Eqn.~\eqref{EqLoading}. 
At this frequency, there are no bulk modes in the linear lattice and edge waves traverse through the lattice. As discussed in 
Sec.~\ref{Sec.Softening}, nonlinear interactions lead to shortening of the bandgap and edge modes do not propagate at high amplitudes. 

Figures~\ref{softSpring_Low1}-(b) display the displacement magnitude in the lattice for 
the low amplitude excitation case 
with $A = 2 \times 10^{-3} $ at two time instants $\tau = 400$ and $\tau = 1000$. 
The color scale has a maximum value $3\times 10^{-3}$ and a minimum value $1.6\times 10^{-3}$. 
We observe edge waves propagating through the lattice in the clockwise direction. Figures~\ref{softSpring_High1}-(d) display 
the magnitude of the displacement vector at each lattice site for the high amplitude force excitation case with $A = 2.0\times 10^{-1}$ at the same time instants.  
The color scale has a maximum value $3\times 10^{-1}$. It is observed that 
energy propagates into the lattice and the amount of energy concentrated on the edge is lower than in the low amplitude case. 
However, note that the waves propagate into the interior only until the amplitude of the edge wave is higher than the threshold required to 
have bulk modes. Note that as the amplitude keeps decreasing, it will reach a value where edge modes are supported. Edge waves at or below 
this threshold keep propagating and the lattice could be seen as a low amplitude 
pass filter for edge waves at this particular excitation frequency.  
\tblue{Thus we observe that, in contrast to the low amplitude case, there is no wave propagation along the boundary. There are
compact zones of energy localization where the displacement is high. These zones are attributed to energy localization 
as a consequence of multiple reflections of bulk waves. Indeed as the dynamics evolves, these localized zones arise at different parts
of the lattice boundary. Note that these are not compactly supported solitons that traverse a boundary.}
\tred{In conclusion, introducing nonlinearity provides  a means to achieve tunability by varying the  wave amplitude. Thus for a given frequency, we illustrated binary behavior: edge waves at one 
amplitude and bulk waves at another amplitude, by careful design of the lattice properties and loading conditions. }

\section{Conclusions}\label{sec:concl}

This work illustrates how localized modes can be induced at the interface or boundaries of both one and two-dimensional lattices. 
In the one-dimensional case, we consider a lattice of point masses connected by alternating springs. We showed that a mode exists 
in the bandgap frequencies and it is localized at the interface between two lattices which are inverted copies of each other. 
We derive explicit expressions for the frequencies of the localized modes for various interface types and their 
associated mode shapes. This localized interface mode can be made tunable by using weakly nonlinear springs at the interface of the two 
masses. We showed that the behavior of the interface mass is equivalent to a Duffing oscillator in the vicinity of this 
interface mode frequency and demonstrated how varying the force amplitude can lead to a frequency shift of the interface mode. By choosing the
parameters carefully, one can control the existence of interface modes and move them from the bandgap to the bulk bands 
by varying the force excitation amplitude. 

In the second part,  we investigate tunability using weakly nonlinear springs in a lattice which supports edge waves. We show how 
the dynamic response of the lattice can be varied from bulk to edge waves at a fixed frequency by varying the excitation amplitude. 
We use an asymptotic analysis to derive dispersion relations for both strain hardening and strain softening springs and demonstrate that 
the optical band can be shifted upward or downward. Finally, numerical simulations are presented to exemplify the theoretical predictions 
and illustrate the tunable nature of our lattices. This work illustrates how exploiting nonlinearities can lead to tunable lattices and mechanical 
structures supporting localized modes at interfaces and boundaries, and it opens the doors for future research in tunable 
engineering structures and devices.

\appendix 

\section{Interface modes}\label{Sec.TransferMatrix}
We seek the frequencies for which the above linear chain admits a localized mode solution at the interface
and derive explicit expressions 
for their corresponding mode shapes. 
Let us consider a finite lattice having $N$ unit cells on either side of the interface, with $N$ large 
enough such that boundary effects are negligible in the dynamics of the interface mass. 
The unit cells are indexed from $p=-N$ to $N$ so that the interface mass lies in the unit cell $p=0$. 
To investigate the dynamics of this lattice in the bandgap frequencies, 
we impose a solution of the form $\bu_p(t) = \bu_p e^{i\Omega \tau}$, for all the lattice sites where $p$ denotes the cell index. 
A similar solution is also imposed on the interface mass. 
To relate the displacements in two neighboring cells $p-1$ and $p$ on the right of the interface ($p>0$), we rewrite the 
governing equations for the masses at the lattice sites $b_{p-1}$
and $a_p$ as
\begin{subequations}
	\begin{align}
	(2 -  \Omega^2) u_{a,p} - (1+\gamma) u_{b,{p}} - (1-\gamma) u_{b,{p-1}} &= 0 \\  
	(2 -  \Omega^2) u_{b,{p-1}} - (1-\gamma) u_{a,{p}} - (1+\gamma) u_{a,{p-1}} &= 0  
	\end{align}
\end{subequations}
Rearranging the terms in the above equation yields a relation between the displacements of adjacent unit cells on the right side of the interface. In nondimensional form, this relation is expressed using a transfer matrix $\bT$ as  
\begin{align}
\begin{pmatrix} u_a \\ u_b\end{pmatrix}_p  &=  
\begin{pmatrix} 
\dfrac{\gamma+1}{\gamma-1} &  \dfrac{2 - \Omega^2}{1 - \gamma} \\
-\dfrac{2 - \Omega^2}{1 - \gamma} & \dfrac{(2-\Omega^2)^2 - (\gamma-1)^2}{1 - \gamma^2}
\end{pmatrix} 
\begin{pmatrix} u_a \\ u_b\end{pmatrix}_{p-1}  \nonumber \\ &= \bT \begin{pmatrix} u_a \\ u_b\end{pmatrix}_{p-1}  . 
\end{align}\label{Eqn.TransferMatrix}
Using the above relation, the displacement at unit cell $p=N$ may be written in terms of the displacement at the interface unit cell ($p=0$), as $\bu_N = \bT^N \bu_0$. Note that the vector $\bu_0$ has components $\bu_0 = (u_{c,0},\;u_{b,0})^T$.

We now solve for the frequencies and corresponding mode shapes at which this chain has localized modes. 
We seek solutions which are localized at the interface and decay away from it, i.e., $\| \bu_N \| \to 0$ as $N$ becomes large. The solution procedure involves seeking 
eigensolutions of the  transfer matrix $\bT$ which satisfy the decay condition. 
For a mode localized at the interface, the displacement should decay away from the interface, i.e., 
$\|\bu_N\| \to 0$ as $N\to \infty$. To make further progress, we use the following
proposition:
	Let $\bT$ be diagonalizable and let  $(\lambda_i,\be_i$) be its eigenvalue-vector pairs. Then, 
	$\| \bT^N \bu\| \to 0$ as $N\to \infty$ with a non trivial solution $\bu\neq\bzero$
	if and only if $\bu$ is in the subspace spanned by the eigenvectors $\be_i$ whose corresponding eigenvalues 
	satisfy $|\lambda_i| < 1$. 
To prove this statement, let us denote by $\bff_i$ the subset of eigenvectors of $\bT$ with associated eigenvalues $|\lambda_i| < 1$ 
and $\bg_j$ the eigenvectors with $|\lambda_j| \geq 1$.
If $u =\sum \alpha_i \bff_i$, 
then $T^N u = \sum\alpha_i {\lambda_i}^N \bff_i$ and hence its norm goes to zero as $N$ increases. 
We prove the `only if' part by contradiction. Assume that $\bu$ is not in the $\bff_i$ subspace as required. We may write 
$\bu = \sum_i\alpha_i\bff_i + \sum_j \beta_j \bg_j$. Then $\bT^N \bu = \sum_i \alpha_i {\lambda_i}^N \bff_i + \sum_j \beta_j {\lambda_j}^N \bg_j $. Since there is a nonzero 
$\beta_j$ by assumption, the norm of this vector does not converge to 0 as $N\to \infty$, which completes the proof. 

Note that the product of the eigenvalues of the transfer matrix
$\bT$ is unity since $\det (\bT ) = 1$. In the bandgap frequencies, the eigenvalues of $\bT$ are real and distinct, 
hence exactly one eigenvalue satisfies $|\lambda_i| < 1$.
The eigenvector corresponding to this eigenvalue is 
\begin{equation}\label{Eqn.SolFreq}
\be = \begin{pmatrix} 
2(\Omega^2-2)^2(1+\gamma)  \\ 
(\Omega^2-2)^2+4\gamma+\Omega\sqrt{(\Omega^2-4)( (\Omega^2-2)^2-4\gamma^2)}
%(\Omega^2-2)^2+4\gamma+\Omega\sqrt{(\Omega^2-2)^2(\Omega^2-4)-4\gamma^2(\Omega^2-4)}
\end{pmatrix} . 
\end{equation}

The proposition above implies that a localized mode arises if the displacement $\bu$ is a scalar multiple by the eigenvector $\be$, 
i.e., $\be = s\bu_0$, with $s$ being a scaling factor and 
$\bu_0 = (u_{c,0}, \; u_{b,0})$ having the displacement components of the unit cell at the interface. 
Let us now derive an expression for $\bu_0$ from the governing equation 
of the interface mass. It may be rewritten as 
\begin{equation}\label{govEinterface}
2\left( 1- \dfrac{\Omega^2}{2(1+\gamma)}\right) u_{c,0} = ( u_{b,0} + u_{b,-1}). 
\end{equation}
Since the localized mode is non-propagating and the lattices on either side of the interface mass are identical, symmetry conditions 
lead to the following relation between the masses adjacent to the interface mass
\begin{equation}
|u_{b,0}| = |u_{b,-1}|. 
\end{equation}
The above condition may be rewritten as $u_{b,-1} = e^{2i\theta} u_{b,0}$. Substituting this into Eqn.~\eqref{govEinterface}, the 
displacement $\bu_0$ may be written as
\begin{equation}
\bu_0 = \begin{pmatrix} e^{i\theta}\cos\theta \\ 1 -  \dfrac{\Omega^2}{2(1+\gamma)} \end{pmatrix} . 
\end{equation}

Note that the chain has bandgaps in the frequency ranges $\Omega \in [\sqrt{2(1-|\gamma|)},\sqrt{2(1+|\gamma|)} ]$ and $\Omega> 2$, 
see Appendix~\ref{Sec.BandInversion} for details. 
Hence, the argument of the square root in Eqn. \eqref{Eqn.SolFreq} is positive when $\Omega$ is in the bandgap frequencies and the 
components of $\be$ are real. The condition $\be  = c\bu_0 $ implies $\theta = n\pi/2, n \in \mathbb{Z}$ 
and $e^{i\theta}\cos\theta \in\{0,1\}$. 
Applying this condition $(e_1/u_{c,0} = e_2/u_{b,0} = c)$ to the two cases separately allows us to solve for the frequencies $\Omega_i$ 
of the localized modes. $\theta = \pi/2$ leads to $\Omega = \sqrt{2}$, while $\theta =0$ leads to the following equation
\begin{equation}\label{Eqn.EMProb}
( \Omega^2-2(1+\gamma))(\Omega^2-2(1-\gamma))(\Omega^2-4) = 4\gamma^2\Omega^2. 
\end{equation}
Note that $\theta=0$ implies  $u_{b,0} = u_{b,1}$. From the transfer matrix expression, we note that 
the mode shape is indeed anti-symmetric about the interface mass. In contrast, $\theta= \pi/2$ results 
leads to $u_{b,0} = -u_{b,-1}$ and $u_{c,0} = 0$. In this case the mode shape is symmetric about the interface mass.  
Equation~\eqref{Eqn.EMProb} leads to the following expressions for the frequencies which support localized solutions
\begin{equation}\label{Eqn.xtra3.Appendix}
\begin{split}
& \gamma < 0:\quad\Omega=\sqrt{ 3- \sqrt{1 + 8 \gamma^2}},\quad\textrm{Anti-symmetric mode} \\
& \gamma > 0:\quad\Omega=\sqrt{2},\;\;\qquad\qquad\quad\quad\textrm{Symmetric mode} \\
& \gamma > 0 :\quad\Omega=\sqrt{ 3+ \sqrt{1 + 8 \gamma^2}},\quad\textrm{Anti-symmetric mode. } 
\end{split}
\end{equation}
Substituting the frequencies $\Omega$ into the eigenvectors in  Eqn. \eqref{Eqn.SolFreq}, 
taking appropriate signs under the square root and checking the condition $\be=c\bu_0$ show
that the first solution is valid when $\gamma<0$, and the other two solutions are valid when $\gamma>0$. 
The displacement components of the interface unit cell for these localized modes are given by 
\begin{equation}\label{Eqn.Sol}
\be = \begin{pmatrix} e_1 \\ e_2  \end{pmatrix} = \begin{pmatrix} u_{c,0} \\ u_{b,0}   \end{pmatrix} . 
\end{equation}
from which the displacement $\bu_p$ of unit cell $p$ can be obtained by using the relation $\bu_p = \bT^p \bu_0$.

\section{Band inversion in linear chain}\label{Sec.BandInversion}

\begin{figure}[hbtp]
	\centering
	\includegraphics[width=0.5\textwidth]{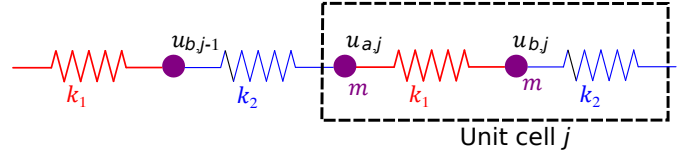}
	\caption{Unit cell of an infinite spring mass chain having alternating spring stiffness $k_1$ and $k_2$. }
	\label{Fig.System}
\end{figure}

We consider a spring mass chain with springs of alternating stiffness $k_1$ and $k_2$ 
connecting identical masses as illustrated in Fig.~\ref{Fig.System}.
The unit cell is chosen as shown by the dashed box in Fig.~\ref{Fig.System}
To normalize the governing equations, we express the spring stiffness as $k_1 = k(1+\gamma)$ and $k_2 = k(1-\gamma)$. 
Introducing non-dimensional time scale  $\tau = t\sqrt{k/m}$, 
the governing equations for the masses in a unit cell may be expressed in non-dimensional form as 
\begin{gather*}
\ddot{u}_{a,j} + 2u_{a,j} - (1+\gamma)u_{b,j} - (1-\gamma)u_{b,j-1}= 0, \\ 
\ddot{u}_{b,j} + 2u_{b,j}- (1+\gamma)u_{a,j} - (1-\gamma)u_{a,j+1}= 0. 
\end{gather*}

We first study the dynamic behavior of the lattice using a dispersion analysis. 
Imposing a plane wave solution of the form 
${\bu}_{j} = (u_{a,j},u_{b,j})=\bA(\mu) e^{i\Omega \tau+i \mu j}$
where $\Omega$ is the frequency and $\mu$ is the non-dimensional wavenumber leads to the following eigenvalue problem 
\begin{multline}\label{Eqn.LinearProb}
\begin{pmatrix}
2-\Omega^2 & -(1+\gamma)-(1-\gamma)e^{-i\mu} \\-(1+\gamma)-(1-\gamma)e^{i\mu} & 2-\Omega^2 
\end{pmatrix}
\begin{pmatrix} 
A_a \\ A_b 
\end{pmatrix} \\
= \Omega^2 \begin{pmatrix} 
A_a \\ A_b 
\end{pmatrix} .
\end{multline}
The eigenvalues lead to two branches with frequencies 
$\Omega=\sqrt{2\pm\sqrt{2+2\gamma^2+2(1- \gamma^2) \cos{\mu}}}$, with the
minus and plus signs for the acoustic and optical bands, respectively. 
The lattice has a bandgap over the frequency range $\Omega \in \left(\sqrt{2(1-|\gamma|)},\sqrt{2(1+|\gamma|)}  \right) $.

%\subsection{Band inversion}

\begin{figure*}[hbtp]
	\centering
	\subfigure[]
	{\includegraphics[width=0.24\textwidth]{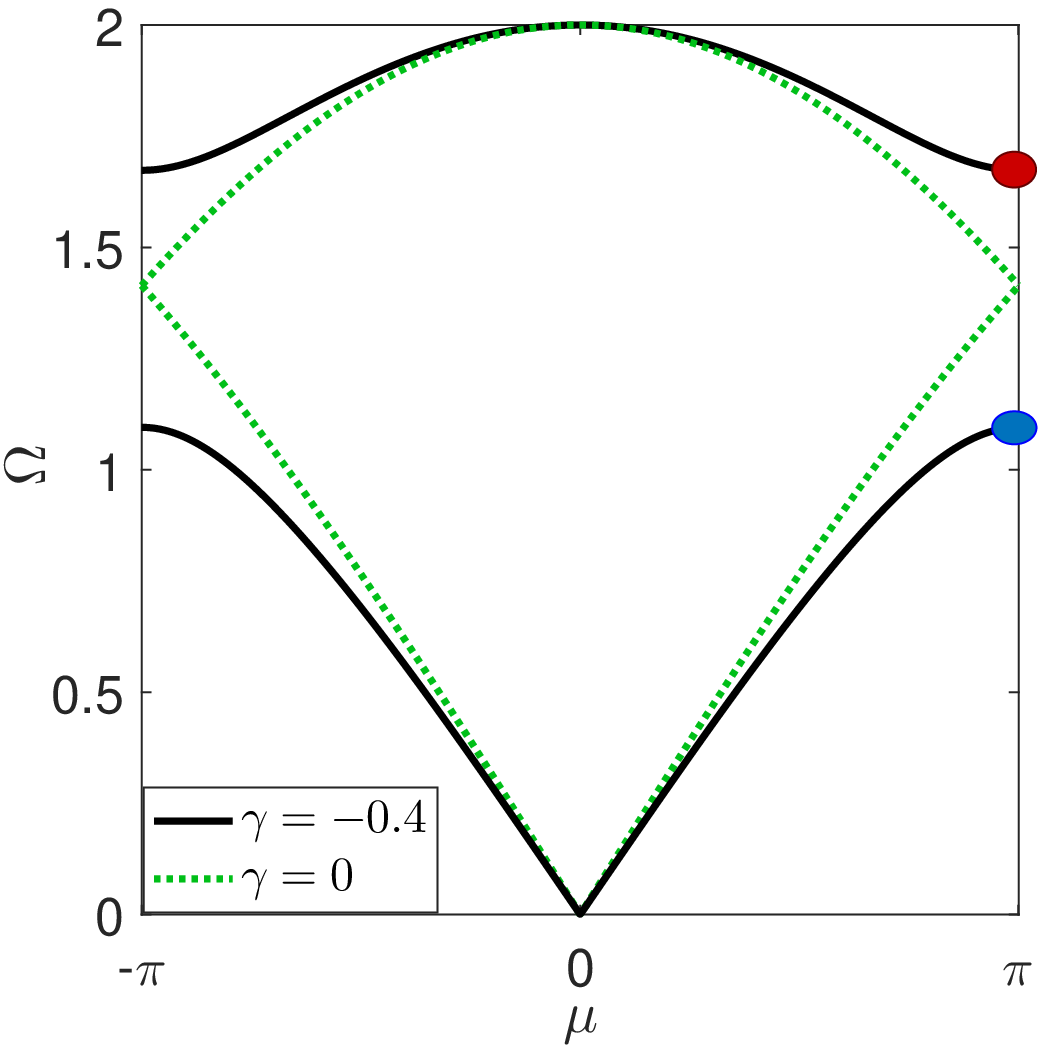}\label{Fig.Disp.1}}
%	\subfigure[]
%	{\includegraphics[width=0.16\textwidth]{Figures/2b.eps}\label{Fig.Disp.3}}
	\subfigure[]
	{\includegraphics[width=0.24\textwidth]{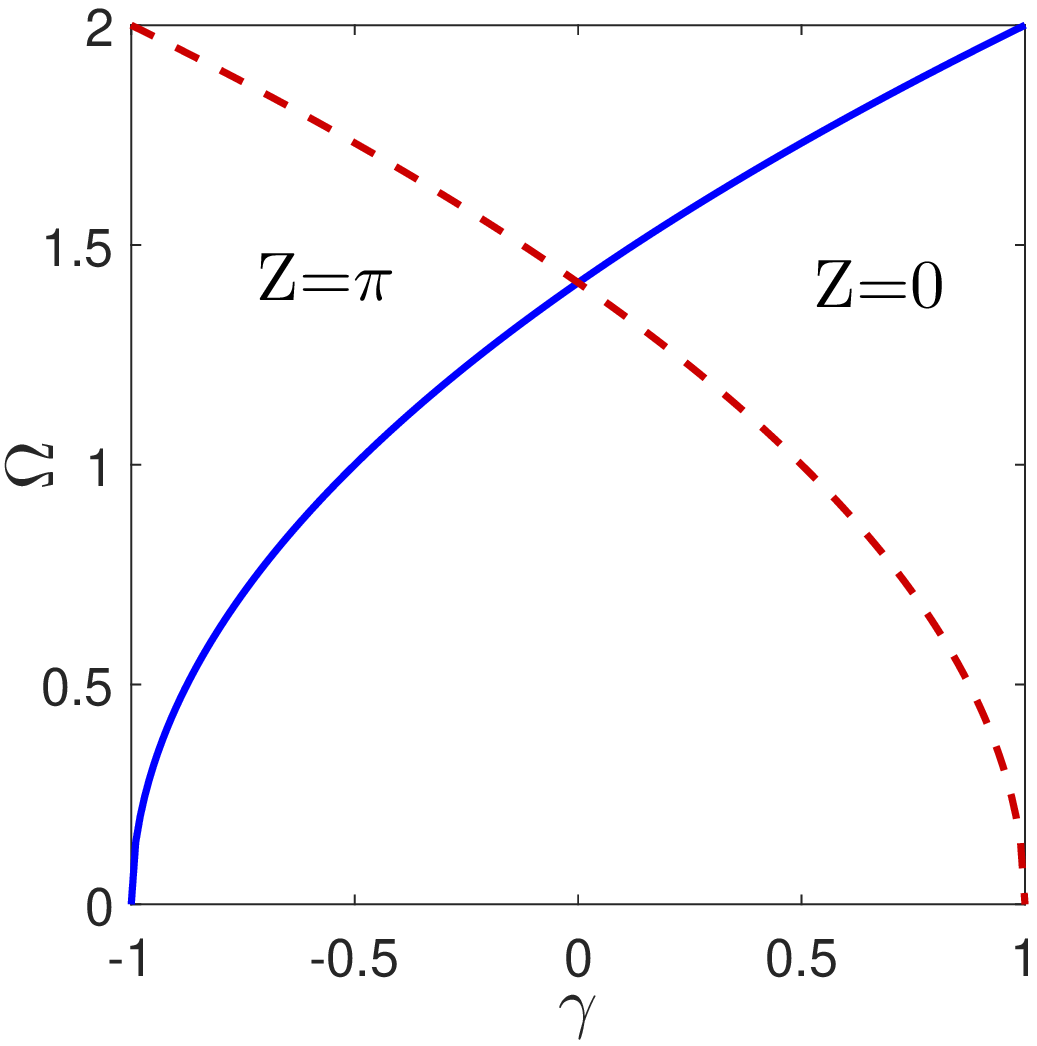}\label{Fig.BGapFreq}}
		\subfigure[]
	{\includegraphics[width=0.24\textwidth]{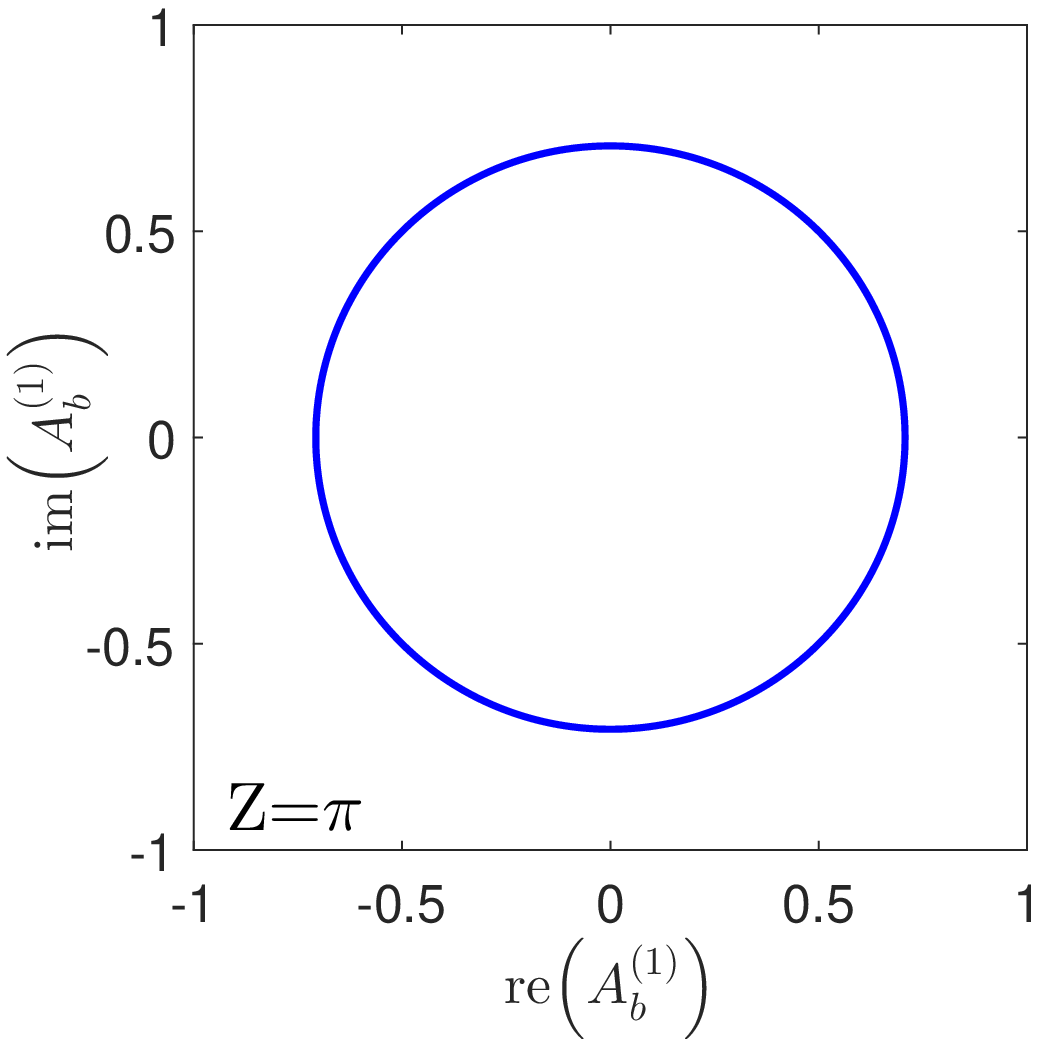}\label{Fig.EigVectorsAndZakPhase.a}}
	\subfigure[]
	{\includegraphics[width=0.24\textwidth]{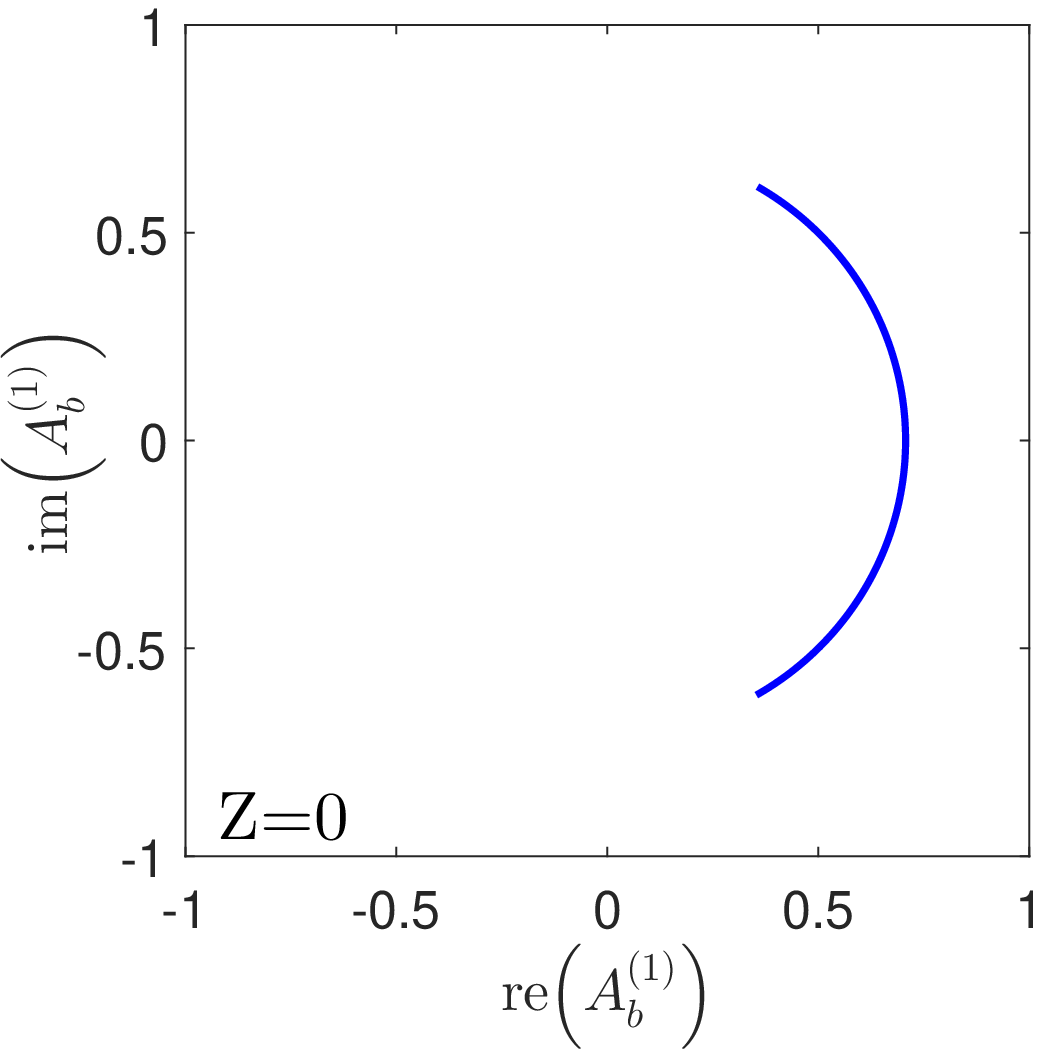}\label{Fig.EigVectorsAndZakPhase.b}}
	\caption{(a) Dispersion relations for lattices with $\gamma=0$ (dotted) and $\gamma=-0.5$ (solid). 
	(b) Limits of the bandgap showing band inversion as $\gamma$ varies. Component $2$ of eigenvector $\bA^{(1)}(\mu)$ 
	as $\mu$ varies from $-\pi$ to $\pi$. (c) $\gamma=-0.5$ has a Zak phase $\pi$ while (d) $\gamma=0.5$ has a zero Zak phase.}	
	\label{Fig.EigVectorsAndZakPhase}
\end{figure*}

Figure~\ref{Fig.Disp.1} displays the dispersion diagrams for stiffness 
parameters $\gamma = 0$  (green dotted curves) and $\gamma = 0.4$ (black solid curves). 
Figure~\ref{Fig.BGapFreq} displays the frequencies bounding the bandgap between the acoustic and optical branches as the stiffness 
parameter $\gamma$ varies. Note that these 
bounding frequencies are at  the wavenumber $\mu = \pi$. 
The frequency on the dashed (red) curve has an eigenvector 
$(A_a,A_b) = (1/\sqrt{2},\; 1/\sqrt{2})^T$ while that on the solid (blue) curve has an eigenvector $(1/\sqrt{2},\; -1/\sqrt{2})^T$ 
for all nonzero $\gamma $ values. Note that the modes get inverted, i.e., the antisymmetric $(1/\sqrt{2},\; -1/\sqrt{2})^T$ 
mode has a higher frequency 
than the symmetric $(1/\sqrt{2},\; 1/\sqrt{2})^T$ mode 
as $\gamma$ increases beyond zero. This phenomenon is called band inversion and has been previously exploited in electronic 
systems \cite{Hasan2010,Bernevig2006,Pankratov1987} and 
continuous acoustic ones \cite{xiao2015geometric} to obtain localized modes. These modes are localized 
at the interface of two lattices: one with $\gamma > 0$ and the other with $\gamma < 0$.

To further shed light on the topological properties of the eigensolutions, 
we examine the eigenvectors of lattices with $\gamma>0$ and $\gamma<0$. In particular, we study how they vary with wavenumber $\mu$
over the first Brillouin zone. 
Observe that the matrix in Eqn.~\eqref{Eqn.LinearProb} gives the same eigenvalues under the transformation 
$\gamma \to -\gamma$ but the eigenvectors are different. Indeed, note that the transformation $\gamma\to -\gamma$ may be
achieved by simply reversing the direction of the lattice basis vector. An alternate way is to simply translate the unit cell by one mass 
to the right or left and relabel the masses appropriately. Both these changes correspond to changes in gauge and they change the eigenvectors,
thereby changing the topology of the vector bundle associated with the solution of the above eigenvalue problem. 
We characterize the topology of this vector bundle using the Zak phase~\cite{zak1989berry} for the bands. 
This quantity is a special case of Berry phase~\cite{zak1989berry,xiao2015geometric} 
to characterize the band topology in $1D$ periodic media. It is given for the band $m$ by
\begin{equation}\label{Eqn.ZakContinuous}
Z=\int_{-\pi}^{\pi}\left[ i (\bA^{(m)})^H(\mu)\cdot\partial_{\mu}\bA^{(m)}(\mu)\right] d\kappa, 
\end{equation}
where $(\bA^{(m)})^H(\mu)$ is the conjugate transpose or Hermitian of the eigenvector $\bA^{(m)}(\mu)$.
For numerical calculations, we use an equivalent discretized form of Eqn. \eqref{Eqn.ZakContinuous} given by~\cite{xiao2015geometric}
\begin{equation}\label{Eqn.ZakDiscrete}
\theta^{Zak}=-\textrm{Im}\sum_{n=-N}^{N-1}\ln\left[\bA_m^H\left({\frac{n}{N}\pi}\right) \cdot\bA_m\left({\frac{n+1}{N}\pi}\right) \right].
\end{equation}
The Zak phase of both the acoustic and optical bands
takes the values $Z = 0$ and $Z=\pi$ for the lattices with $\gamma > 0$ and $\gamma < 0$, respectively.  
Indeed, it should be noted that since the Zak phase is not gauge invariant~\cite{atala2013direct}, 
the choice of coordinate reference and unit cell must remain the same for computing this quantity. 

To understand the meaning of the Zak phase, we show the behavior of the 
acoustic mode eigenvector for both $\gamma>0$ and $\gamma < 0$ lattices. 
For consistent representation, a gauge  is fixed  such 
that the eigenvector has magnitude $1$ and its first component is real and positive, i.e., at zero angle in the complex plane.
The second component $A_b^{(1)}$ of the eigenvector is displayed in the complex plane for $\mu$ varying from $-\pi$ to $\pi$, see Figs. \ref{Fig.EigVectorsAndZakPhase.a} and \ref{Fig.EigVectorsAndZakPhase.b}. This component of the eigenvector will form a loop as the wavenumber
$\mu$ is varied from $-\pi$ to $\pi$. 
When $\gamma>0$, this eigenvector loop does not enclose the origin and it leads to a Zak phase equal to $0$. On the other hand, 
the acoustic band of a lattice with $\gamma<0$ has a Zak phase of $Z=\pi$ and its eigenvector loop $A_b(\mu)$ encloses the origin.

\section{Effective stiffness of interface mass}\label{Sec:ReducedSoln}

We consider a finite lattice with an interface and where the masses at both ends are fixed. Using the transfer matrix relations, 
we derived the following expression in Sec.~\ref{Sec.TransferMatrix} for the equivalent stiffness of the interface mass 
\begin{equation*}
\left[ 2(1+\gamma) \left( 1 - \dfrac{S_{21}}{S_{11}}\right)  - \Omega^2 \right] u_{c,0}= f,  
\end{equation*}
where $\bS = \bT^{-N}$. Let us now derive an explicit expression for the terms of the matrix $\bS$ which appear in the above expression. 
Let us assume that $\bT^{-1}$ is diagonalizable. This assumption is verified later by examining its eigenvectors. We now use 
the following result
from linear algebra~\cite{hoffmanlinear}: there exists a 
unique decomposition $\bT^{-1} = \bU \bD \bU^{-1}$, where $\bD$ is a diagonal matrix having the eigenvalues of $\bT^{-1}$ and $\bU$ is a matrix
whose columns are the corresponding eigenvectors of $\bT^{-1}$. We determine this decomposition by solving for the eigenvectors of $\bT^{-1}$, 
which then leads to the following expression 
for $\bS$
\begin{align}
\bS &= \bT^{-N} = \bU \bD^N \bU^{-1} = [\bU]  
\begin{pmatrix}\lambda_1^N & 0 \\ 0 & \lambda_2^N \end{pmatrix} 
  [\bU^{-1}],
\end{align}
where 
\begin{widetext}
\begin{gather*}
\lambda_{1,2} =  \dfrac{-2-2\gamma^2+(2-\Omega^2)^2 \pm 
\Omega \sqrt{(\Omega^2-4)(-4\gamma^2 + (2-\Omega^2)^2)}}{2(1-\gamma^2)}, 
\\
[\bU] = \begin{pmatrix} 
\dfrac{4\gamma + (2-\Omega^2)^2 + \Omega\sqrt{(\Omega^2-4)(-4\gamma^2 + (2-\Omega^2)^2)} }{2(1+\gamma)(2-\Omega^2)} &  
\dfrac{4\gamma + (2-\Omega^2)^2 - \Omega\sqrt{(\Omega^2-4)(-4\gamma^2 + (2-\Omega^2)^2)} }{2(1+\gamma)(2-\Omega^2)}  \\ 
1& 1
\end{pmatrix}. 
\end{gather*}
\end{widetext}
Note that the term within the square root in the eigenvectors is positive for all frequencies $\Omega$ in the bandgaps and thus the 
two eigenvectors are distinct. These two distinct eigenvectors span the vector space $\mathbb{R}^2$ and hence $\bT^{-1}$ is 
diagonalizable~\cite{hoffmanlinear}, which verifies the assertion made earlier. 

%%%%%%%%%%%%%%%%%%%%%%%%%%%%%%%%%%%%%%%%%%%%%%%%%%%%%%%%%%%%%%%%%%%%%%%%%%%%%%%%%%%%%%%%%%%%%%%%%%%%%%%%%%%%%%%%%%%%%%%%%%
%%%%%%%%%%%%%%%%%%%%%%%%%%%%%%%%%%%%%%%%%%%%%%%%%%%%%%%%%%%%%%%%%%%%%%%%%%%%%%%%%%%%%%%%%%%%%%%%%%%%%%%%%%%%%%%%%%%%%%%%%%
%%%%%%%%%%%%%%%%%%%%%%%%%%%%%%%%%%%%%%%%%%%%%%%%%%%%%%%%%%%%%%%%%%%%%%%%%%%%%%%%%%%%%%%%%%%%%%%%%%%%%%%%%%%%%%%%%%%%%%%%%%
\section{Asymptotic analysis procedure}\label{asymptoticTheory}
To seek the dispersion relation of the lattice, the governing equation for a lattice unit cell $j$ are written in matrix form as 
\begin{equation}
\bM \ddot{\bu}_{j} + \sum_p \bK_{p} \bu_{p}  + f_{NL}\left( \bu_j,\bu_p   \right) = \bzero.  
\end{equation}
The displacement and frequency are solved using the method of multiple scales, having the following asymptotic expansions for displacement
$\bu$ and time $t$
\begin{equation}
\bar\bu_j=\bu_j^{(0)}+\epsilon\bu_j^{(1)}+O(\epsilon^2), \;\;
t=\omega \tau  =  ( \omega_0+\epsilon\omega_1+O(\epsilon^2) ) \tau.
\end{equation}
Assuming $\bu_j$ is harmonic with frequency $\omega$, and 
substituting the above equations into the governing equations (Newton's laws) 
yields the following equations for the various orders of $\epsilon$:
\begin{align}\label{govE0}
&\epsilon^0: \; \;\omega_0^2 \bM \dfrac{d^2 \bu^{(0)}_{m}}{d \tau^2} + \sum_{p} \bK_{p} \bu^{(0)}_{m+p} = 0,  \\
&\epsilon^1: \; \; \omega^2_0 \bM \dfrac{d^2 \bu^{(1)}_{m}}{d \tau^2} + \sum_{p} \bK_{p} \bu_{m+p}^{(1)} =
-2 \omega_0 \omega_1 \bM \dfrac{d^2 \bu_{m}^{(0)}}{d\tau^2}  \nonumber \\
&\qquad \qquad \qquad \qquad \qquad 
-\sum_{p} f_{NL} \left( \bu^{(0)}_{m} , \bu^{(0)}_{m+p} \right).\label{govEqn_1}
\end{align}
The solution of the above equations yields the first order plane waves and their amplitude dependent dispersion relations.

The zeroth order equation is linear and is solved 
using the Floquet Bloch theory. We impose a traveling wave solution of the form
$\bu^{(0)}_p(\tau) = \bz_m e^{i\bmu \cdot \bx_{p}} e^{i\tau}$,
where $\bz$ is the eigenvector associated with a wave with wavevector $\bmu$, $\bx_{p}$ is the spatial location 
of the center of a unit cell with index $p$ and $\bu^{(0)}_p$ is the vector with components having the generalized displacement 
of unit cell $p$. 
Substituting this expression into the system of governing equations for a unit cell in Eqn.~\eqref{govE0} leads to
the eigenvalue problem
$\sum_{p} \bK_{p} e^{i\bmu \cdot \bx_p} \bz=\omega_0^2 \bM \bz $ 
for a fixed wavevector $\bmu$ in the reciprocal lattice space. 
Its solution 
yields the dispersion surface $\omega = \omega(\bmu)$ of the zeroth order linear system. 
Then the zeroth order displacement 
of a cell $p$ due to the $m$-th wave mode, with eigenvector $\bz_m$, may be written as 
\begin{equation}\label{zeroSoln2}
\bu^{(0)}_{p}(\tau)  =  \dfrac{A_0}{2}  \left( \bz_{m}(\bmu) e^{i\bmu\cdot \bx_p} e^{i\tau} + c.c.\right),
\end{equation}
where $c.c.$ denotes the complex conjugate, $A_0$ is the wave amplitude and $\bz_{m}(\bmu)$ is the $m$-th wave mode 
at the wavevector $\bmu$. 
The eigenvector ${\bz_m}(\bmu) $  is normalized so that the maximum absolute value of any 
component is $1$. Thus the maximum displacement of any mass in the lattice is $A_0$.

In contrast with linear media, the dispersion behavior of our  nonlinear 
lattice will depend on the amplitude $A_0$ of the wave mode and we determine the 
first order correction in the dispersion relation as a function of this amplitude. 
To this end, the first order equation is solved to get a
correction due to the nonlinear terms. 
Substituting Eqn.~\eqref{zeroSoln2} into the first order equation Eqn.~\eqref{govE0} leads to the following equation corresponding to 
the $j$-th wave mode
\begin{multline}\label{Eqn.SecularTerms}
\omega_{0}^2 \bM \dfrac{d^2 \bu^{(1)}_m}{d \tau^2} + \sum_{p} \bK_{p,m} \bu_{p+m}^{(1)}  =  
\omega_0 \omega_1 A_0 \bM \bu^{(0)}_m e^{i\tau} \\
- \sum_p f_{NL}(\bu_m^{(0)} , \bu^{(0)}_{m+p}) = \bF(\bmu).
\end{multline}
Note that the linear part of Eqn.~\eqref{Eqn.SecularTerms} (terms on the left) 
is identical to the $\epsilon^0$ order equation (Eqn.~\eqref{govE0}).
The term $\bF(\bmu)$ on the right hand side is the additional forcing term in the $\epsilon^1$ order equation
due to nonlinear effects. 
The component of $\bF$ along $\bu^{0}_j$ is identified as a secular term and it 
should vanish for the $\epsilon^1$ solution to be 
bounded. This condition may be written as
\begin{equation}
\left( \bu^{0}_{j} , \bF(\bmu )\right) = \int_0^{2\pi} (\bu^{0}_j)^H \bF(\bmu) d\tau = 0. 
\end{equation}
Substituting terms from Eqn.~\eqref{Eqn.SecularTerms} into $\bF$,  the above condition leads to an equation for the 
first order frequency correction $\omega_{1}$. For the $j$-th mode, solving this equation gives
\begin{equation}\label{Eqn.Omeg1}
\omega_{1}(A_0 , \bmu)   =  \dfrac{ {\bu^{(0)H}_{m}} }{2 \pi\omega_{0,j} A_0 {\bu^{(0)H}_{m}} \bM {\bu^{(0)}_{m}} } \int_0^{2\pi} \sum_p  f_{NL} e^{-i\tau} d \tau. 
\end{equation}
Note that since $\bu^{(0)}$ is a periodic function of 
time $\tau$ with period $2\pi$, the nonlinear forcing function $f_{NL}$ is also periodic in $\tau$. 
The nonlinear forces also depend on the amplitude $A_0$ of the zeroth order wave mode. 

We now address two technical points which ensure uniqueness of the above expression for $\omega_{1,j}$. The first one is when there
are repeated eigenvalues and the second one is about the invariance of the correction $\omega_1$ to the 
scaling of eigenvectors by $e^{i\theta}$. 
A procedure is outlined to address the case of repeated eigenvalues in a systematic way which results in a unique and well defined 
value of the frequency correction. Let us consider a wavevector $\bmu$ at which there are $p$ repeated eigenvalues $\omega_0$ and the
corresponding eigenvectors are $\{\bv_i:1 \le i\le p\}$. A linear combination of any of these $p$ eigenvectors is also a valid
eigenvector and these eigenvectors define a vector subspace. 
However, note that $\bc_1$ in Eqn.~\eqref{Eqn.Omeg1} depends on the eigenvectors in a nonlinear way and hence its value
will depend on the choice of eigenvectors from this vector subspace.
As an illustrative example, consider the eigenvalues and eigenvectors of the identity $\bI_2$ matrix. It has a repeated eigenvalue $1$ and 
the corresponding eigenvectors are non-unique. We consider two sets of eigenvectors $(\bv_1,\bv_2)$ and $(\bw_1,\bw_2)$:
\begin{gather*}
\bv_1 = \begin{pmatrix}1 \\ 0\end{pmatrix}e^{i\tau}, \;\; 
\bv_2 = \begin{pmatrix}0 \\ 1\end{pmatrix}e^{i\tau}, \\
\bw_1 = \begin{pmatrix}1/\sqrt{2} \\ 1/\sqrt{2}\end{pmatrix}e^{i\tau}, \;\; 
\bw_2 = \begin{pmatrix}1/\sqrt{2} \\ -1/\sqrt{2} \end{pmatrix}e^{i\tau}. 
\end{gather*}
Solving for $\omega_1$ gives different values for the $\bv_i$ and  $\bw_{i}$ sets as $\bc_1$ has a nonlinear dependence on the components. 
To resolve this anomaly, we remark here that the eigenvalue correction corresponds to waves propagating at specific prescribed amplitude $A_0$.
Note that this amplitude is prescribed on distinct eigenvectors in a mutually exclusive manner. For example, if we prescribe 
the zeroth order solution displacement on masses $(i,j)$, the eigenvectors $(\bu_1,\bu_2)$ having common eigenvalue 
should satisfy the constraint 
\begin{equation}\label{eq1}
v_1(i) = \alpha A_0, \;\;v_1(j) = 0, \;\;\;v_2(i) = 0, \;\;v_2(j) = \beta A_0, \;\;\;
\end{equation}
where $\alpha$ and $\beta$ are normalizing constants such that the maximum magnitude of any component is unity 
(i.e., $\|\bv_1\|_{\infty} = \|\bv_2\|_{\infty} = 1$).  

Extending the above observation to the general case having $p$ common eigenvalues with eigenvectors, 
we present  the following approach to get a set of transformed eigenvectors which obey a generalized form of the 
constraint in Eqn.~\eqref{eq1}. For each eigenvector $\bu_i$ in this set of repeated eigenvalues, we find a corresponding transformed 
eigenvector $\bv_i$ by setting $p-1$ components of $\bu_i$ to zero. These $p-1$ components are simply chosen to be those which have
the highest magnitude. Note that if the indices of these components coincide with those chosen for another $i$ 
distinct from this set of repeated of repeated eigenvalues, then a different set of 
indices are chosen. This procedure ensures that we are enforcing distinct components to zero value to get the orthogonal modes. 

The second point about the invariance of $\omega_{1,j}$ to the specific choice of gauge factor $e^{i\theta}$ is explained by noting 
that only the component corresponding to $e^{i\tau}$ in $f_{NL}$ is relevant to the computation of $\bc_1$ and the 
contribution of all other components is zero due to orthogonality. Hence, $\bc_1$ depends linearly on $e^{i\theta}$ and 
multiplying by $\bu^H$, which has a factor $e^{-i\theta}$ ensures that the resulting final expression is independent of $\theta$.

\section*{Acknowledgments}
\label{sec:Acknowledgments}

The authors are indebted to the US Army Research Office (Grant number W911NF1210460),
the US Air Force Office of Scientific Research (Grant number FA9550-13-1-0122) and the National Science Foundation (Grant number 1332862) for financial support. 
\bibliographystyle{unsrt}
\bibliography{paper}

\end{document}